\documentclass[twocolumn,5p,sort&compress]{elsarticle}
\pdfoutput=1
\usepackage{url}
\usepackage{tikz}
\usepackage{comment}
\usepackage{framed,graphicx}
\usepackage{multirow}
\usepackage{rotating}
\usepackage{amsmath}
\usepackage{bigstrut}
\usepackage{color}
\usepackage{graphics}
\usepackage{eqparbox}
\usepackage{colortbl}
\usepackage{paralist}
\usepackage{soul}
\usepackage{xcolor}
\usepackage{caption}
\usepackage{mathptmx} 
\usepackage{listings}
\usepackage{courier}
\usepackage{balance}
\usepackage{picture}
\usepackage{algorithm}
\usepackage{algorithmicx}
\usepackage{algpseudocode}
\usepackage[export]{adjustbox}
\definecolor{lightgray}{gray}{0.8}
\definecolor{darkgray}{gray}{0.6}
\definecolor{lavenderpink}{rgb}{0.98, 0.68, 0.82}
\definecolor{celadon}{rgb}{0.67, 0.88, 0.69}

\definecolor{Gray}{gray}{0.95}
\definecolor{LightGray}{gray}{0.975}

\newcommand{\tion}[1]{\ref{sect:#1}}
\newcommand{\eq}[1]{Equation~\ref{eq:#1}}
\newcommand{\fig}[1]{Figure~\ref{fig:#1}}
\newcommand{\bi}{\begin{itemize}}
\newcommand{\ei}{\end{itemize}}
\newcommand{\be}{\begin{enumerate}}
\newcommand{\ee}{\end{enumerate}}
\usepackage{enumitem}
\setitemize{noitemsep,topsep=0pt,parsep=0pt,partopsep=0pt}

  \biboptions{sort&compress}

\definecolor{steel}{rgb}{.11, .11, .7}
\definecolor{Gray}{rgb}{0.88,1,1}
\definecolor{Gray}{gray}{0.85}
\usepackage[framed]{ntheorem}
\usetikzlibrary{shadows}
\theoremclass{Lesson}
\theoremstyle{break}

\tikzstyle{thmbox} = [rectangle, rounded corners, draw=black,
fill=Gray!40,  drop shadow={fill=black, opacity=1}]

\newshadedtheorem{lesson}{Result}

\usepackage{listings}
\definecolor{MyDarkBlue}{rgb}{0,0.08,0.45} 
\begin{document}
\begin{frontmatter}

\title{What is  Wrong with
Topic Modeling?\\ (and How to Fix it Using Search-based Software Engineering)}

\author{Amritanshu Agrawal\corref{cor1}}

\cortext[cor1]{Corresponding author: Tel: +1-919-637-0412 (Amritanshu)}

\ead{aagrawa8@ncsu.edu}
\ead[url]{https://amritag.wixsite.com/amrit}

\author{Wei Fu\corref{cor1}}
\ead{wfu@ncsu.edu}

\author{Tim Menzies\corref{cor1}}
\ead{tim.menzies@gmail.com}
\address{Department of Computer Science, North Carolina State University, Raleigh, NC, USA}

\begin{abstract}

\noindent \textbf{Context:} Topic modeling finds
  human-readable structures in unstructured textual data. A
  widely used topic modeling technique is Latent Dirichlet allocation. When
  running on different datasets, LDA suffers from ``order effects'', i.e., different topics are generated if the order of training data is shuffled.
  Such order effects introduce a
 systematic error for any study. This error can relate to misleading results;
  specifically, inaccurate topic descriptions and a reduction in the efficacy of
  text mining classification results.

 \noindent
\textbf{Objective:} To provide a method in which distributions
generated by LDA are more stable and can be used for further analysis.

 \noindent
\textbf{Method:} We use LDADE, a search-based software engineering tool which uses Differential Evolution (DE) to tune the LDA's parameters.
LDADE is evaluated on data from a programmer
information exchange site (Stackoverflow), title and abstract text of thousands
of Software Engineering (SE) papers, and software defect reports from NASA. Results were collected
across different implementations of LDA (Python+Scikit-Learn, Scala+Spark) across
Linux platform and for different kinds of LDAs (VEM, Gibbs sampling). 
Results were scored via topic stability and text mining classification accuracy.

\noindent
\textbf{Results:}
In all treatments:
   (i)~standard LDA exhibits very large topic instability;
  (ii)~LDADE's tunings dramatically reduce cluster instability; 
  (iii)~LDADE also leads to  improved performances for supervised as well as unsupervised learning.

\noindent
\textbf{Conclusion:}
  Due to topic instability,
  using standard LDA with its ``off-the-shelf'' settings should now be depreciated.
  Also, in future,
we should require
SE
papers that use
LDA
to
test and (if needed) mitigate LDA topic instability.
  Finally, LDADE is a candidate technology for effectively and efficiently reducing that instability.  
\end{abstract}


\end{frontmatter}

\section{Introduction}
\label{sect:intro}

The current great challenge in software analytics
is understanding unstructured data. As shown in Figure~\ref{fig:
data}, most of the planet's 1600 Exabytes of data does not appear
in structured sources (databases, etc)~\cite{nadkarni2014structured}.
Mostly the data is of {\em unstructured} form, often in free text, and found
in word processing files, slide presentations, comments, etc.

Such unstructured data does not have a pre-defined data model and
is typically text-heavy. Finding insights among unstructured text
is  difficult unless we can search, characterize, and classify the
textual data in a meaningful way. One of the common techniques for
finding related topics within unstructured text (an area called
topic modeling) is Latent Dirichlet allocation (LDA)~\cite{blei2003latent}.

This paper explores systematic errors in LDA analysis.
LDA is a non-deterministic algorithm since its internal weights are updated via a stochastic sampling process (described
later in this paper).
We show in this paper that this non-determinism means that the  topics generated by LDA on
SE data are subject to order effects, i.e., different input orderings
can lead to different topics.
Such instability can:
\begin{figure}[!b]
  \captionsetup{justification=centering}
  \includegraphics[width=\linewidth]{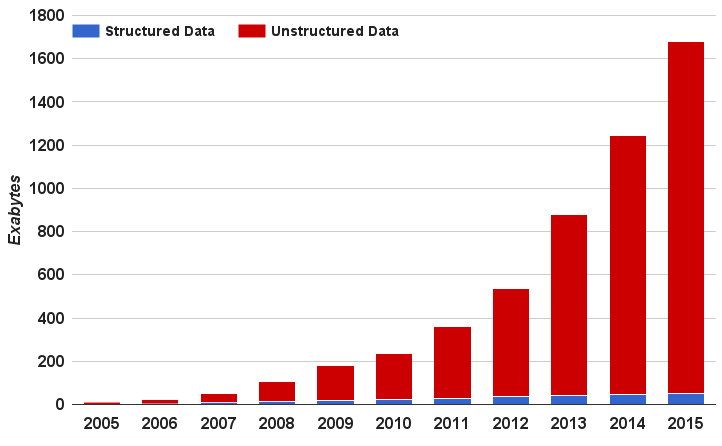}
  \caption{Data Growth 2005-2015. From~\cite{nadkarni2014structured}.}
  \label{fig: data}
\end{figure}

\bi
\item
  Confuse users when they see different topics each time
the algorithm is re-run.
\item
  Reduce the efficacy of text mining classifiers that rely on LDA to
  generate their input training data.
\ei
To fix this problem,
we propose LDADE: a  combination of LDA and a search-based optimizer (differential evolution, or DE)~\cite{storn1997differential})
that automatically tunes LDA's \mbox{$<k$, $\alpha$, $\beta>$} parameters. We prefer LDADE to other methods for two reasons, 1) LDADE is orders of magnitude faster and 2) other methods do not address the problem of order effects (for evidence on this second point,
see our  comparison against the LDA-GA method used by Panichella et al.~\cite{panichella2013effectively}, in the results section).

This paper tests LDADE 
by applying text mining to
three data sets: (a)~Data from a programmer information exchange site (Stackoverflow), (b)~Title and abstract text of 15121 SE papers (Citemap) and (c)~Software defect reports from NASA (Pits).
    
\noindent
Using these datasets, we explore these research questions:  
   
\bi

\item \textbf{RQ1}: \textbf{Are the default settings of LDA incorrect?} We will show that using the default settings of LDA for SE data can lead to systematic errors since stability scores start to drop after $n=5$ terms per topic.
    \item \textbf{RQ2}: \textbf{Does LDADE improve the stability scores?} LDADE dramatically improves stability scores using
      the parameters found automatically by DE. 
    \item \textbf{RQ3}: \textbf{Does LDADE improve text mining classification accuracy?} 
      Our experiments shows that LDADE  also improves classification accuracy.
    \item \textbf{RQ4}: \textbf{Do different data sets
      need different configurations to make LDA stable?} LDADE finds different ``best'' parameter settings for different data sets. Hence reusing tunings  suggested  by  any other  previous study  for any dataset is \underline{{\em not}} recommended. Instead,  it is better to
      use  automatic  tuning  methods  to find the best tuning parameters for the current data set.
      \item \textbf{RQ5}: \textbf{Are our findings consistent when using different kinds of LDA or with different implementations?} Results were collected across different implementations of LDA across Linux platform and for different kinds of LDAs. Across all these implementations,
        the same effect holds: (a)~standard LDA suffers from order effect and topic instability
        and (b)~LDADE can reduce that instability.
    \item \textbf{RQ6}: \textbf{Is tuning easy?} We show that, measured
      in the terms of the internal search space of the optimizer,
      tuning LDA is much simpler than standard optimization methods.
    \item \textbf{RQ7}: \textbf{Is tuning extremely slow?}
      The advantages of  LDADE come at some cost:
      tuning with DE makes LDA three to five times slower.
      While this is definitely more than not using LDA, this may not be an arduous increase
      given modern cloud computing environments. 
    \item \textbf{RQ8}: \textbf{How better LDADE is compared against other methods for tuning LDA?}
    Our analysis confirms that LDADE is better for LDA tuning than (a)~random search and (b)~the genetic algorithm approach used in prior work~\cite{panichella2013effectively}.
       
    \item \textbf{RQ9}: \textbf{Should topic modeling be used ``off-the-shelf'' with their default tunings?}
      Based on these findings, our answer to this question is an emphatic ``no''. We can see little reason to use ``off-the-shelf'' LDA for any kind of SE data mining applications.
      \ei

The rest of this paper is structured as follows.
Section \tion{motivate} argues that stabilizing the topics generated by LDA is important for several reasons.
Related work is reviewed in Section \tion{related} and the methods of this paper are discussed in Section \tion{evaluation}.
We have answered above research questions in
Section \tion{results}. This is followed by a discussion on the validity of our results and a section
describing our conclusions.
Note that the main conclusion
of this paper is that, henceforth,
we should require
SE
papers that use
LDA
to
test and (if needed) mitigate LDA topic instability.

\section{Motivation}
\label{sect:motivate}

\begin{table}[!b]
\renewcommand\arraystretch{1.0}
\begin{center}
\scriptsize
\caption{Top SE venues that published on Topic Modeling from 2009 to 2016.}
\label{tab:venues}
\begin{tabular}{|c|l|c|}
\hline
\begin{tabular}[c]{@{}c@{}}\textbf{Venue}\end{tabular} & \begin{tabular}[c]{@{}c@{}}\textbf{Full Name}\end{tabular}     & \multicolumn{1}{l|}{\textbf{Count}} \\ \hline
ICSE & International Conference on Software Engineering  & 4  \\ \hline
\begin{tabular}[c]{@{}c@{}}CSMR-\\WCRE\\ / SANER \end{tabular}   & \begin{tabular}[c]{@{}l@{}}International Conference on Software Maintenance,\\ Reengineering, and Reverse Engineering / International \\Conference on Software Analysis, Evolution, and \\Reengineering\end{tabular} & 3  \\ \hline
\begin{tabular}[c]{@{}c@{}}ICSM\\ / ICSME\end{tabular}   & \begin{tabular}[c]{@{}l@{}}International Conference on Software Maintenance / \\International Conference on Software Maintenance and \\Evolution \end{tabular}  & 3\\ \hline
ICPC & International Conference on Program Comprehension  & 4  \\ \hline
ASE      & \begin{tabular}[c]{@{}l@{}}International Conference on Automated Software\\ Engineering  \end{tabular} & 3  \\ \hline
ISSRE      & \begin{tabular}[c]{@{}l@{}}International Symposium on Software Reliability\\ Engineering \end{tabular} & 2  \\ \hline
MSR      & \begin{tabular}[c]{@{}l@{}}International Working Conference on Mining \\Software Repositories \end{tabular} & 8  \\ \hline
OOPSLA  & \begin{tabular}[c]{@{}l@{}}International Conference on Object-Oriented \\Programming, Systems, Languages, and Applications \end{tabular}  & 1 \\ \hline
FSE/ESEC  &  \begin{tabular}[c]{@{}l@{}}International Symposium on the Foundations of Software \\Engineering / European Software Engineering Conference \end{tabular}  & 1                                    \\ \hline
TSE                                 & IEEE Transaction on Software Engineering   & 1                                     \\ \hline
IST                                 & Information and Software Technology                                 & 3                                       \\ \hline
SCP                                 & Science of Computer Programming                               & 2                                       \\ \hline
ESE                                 & Empirical Software Engineering                         & 4                                       \\ \hline
\end{tabular}
\end{center}
\end{table}

\renewcommand\arraystretch{1.2}
\begin{table*}[!t]
\scriptsize
\centering
\caption{A sample of the recent literature on using topic modeling in SE. Sorted by number of citations (in column3). 
For details on how this table was generated, see Section \tion{solutions}.}
\label{tbl:survey2}
    \begin{tabular}{|c@{~}|c@{~}|c@{~}|c@{~}|c@{~}|c@{~}|c@{~}|p{4.95cm}@{~}|p{2.75cm}|c@{~}|}
        \hline 
        \begin{tabular}[c]{@{}c@{}}\textbf{REF}\end{tabular} & \textbf{Year} & \textbf{Citations} & \textbf{Venues} & \begin{tabular}[c]{@{}c@{}}\textbf{Mentions} \\\textbf{instability} \\\textbf{in LDA?} \end{tabular} &\begin{tabular}[c]{@{}c@{}} \textbf{Uses} \\\textbf{Default} \\\textbf{Parameters}\end{tabular}&\begin{tabular}[c]{@{}c@{}} \textbf{Does} \\\textbf{tuning?}\end{tabular} & \multicolumn{1}{c|}{Conclusion}  & \multicolumn{1}{c|}{Tasks / Use cases} & \begin{tabular}[c]{@{}c@{}}\textbf{Unsupervised} \\\textbf{or Supervised} \end{tabular} \\ \hline
        \cite{rao2011retrieval} & 2011 & 112 & WCRE & Y & Y & N & \begin{tabular}[c]{@{}l@{}}Explored Configurations without any explanation. \end{tabular} & \begin{tabular}[c]{@{}l@{}}Bug Localisation\end{tabular} & Unsupervised \\ [0.5ex]\hline
        \cite{oliveto2010equivalence} & 2010 & 108 &MSR& Y & Y & N & \begin{tabular}[c]{@{}l@{}}Explored Configurations without any explanation.\\ Reported their results using multiple experiments.\end{tabular}& \begin{tabular}[c]{@{}l@{}}Traceability Link recovery\end{tabular} & Unsupervised\\ [0.5ex]\hline
        \cite{barua2014developers} &2014& 96 & ESE & Y & Y & N  & \begin{tabular}[c]{@{}l@{}}Explored Configurations without any explanation.\\ Choosing right set of parameters is a difficult task.\end{tabular}& \begin{tabular}[c]{@{}l@{}}Stackoverflow Q\&A data \\ analysis\end{tabular} & Unsupervised\\ [0.5ex]
        \hline
        \cite{panichella2013effectively} & 2013&75&ICSE & Y & Y & Y  & \begin{tabular}[c]{@{}l@{}}Uses GA to tune parameters. They determine\\ the near-optimal configuration for LDA in the \\context of only some important SE tasks.\end{tabular}& \begin{tabular}[c]{@{}l@{}}Finding near-optimal \\ configurations\end{tabular} & Semi-Supervised \\ [0.5ex]
        \hline
        \cite{galvis2013analysis} &2013& 61 &ICSE& Y & Y & N  & \begin{tabular}[c]{@{}l@{}}Explored Configurations without any explanation.\end{tabular}& \begin{tabular}[c]{@{}l@{}}Software Requirements \\ Analysis\end{tabular} & Unsupervised \\ [0.5ex]
        \hline
        \cite{hindle2011automated} &2011& 52 & MSR & Y & Y & N  & \begin{tabular}[c]{@{}l@{}}They validated the topic labelling techniques \\using multiple experiments.\end{tabular}& \begin{tabular}[c]{@{}l@{}}Software Artifacts Analysis\end{tabular} & Unsupervised \\ [0.5ex]
        \hline
        \cite{guzman2014users} & 2014 & 44 &RE& Y & Y & N  & \begin{tabular}[c]{@{}l@{}}Explored Configurations without any explanation.\end{tabular}& \begin{tabular}[c]{@{}l@{}}Requirements Engineering\end{tabular} & Unsupervised \\ [0.5ex]
        \hline
        \cite{thomas2011mining} &2011& 44 &ICSE & Y & Y & N  & \begin{tabular}[c]{@{}l@{}}Open issue to choose optimal parameters.\end{tabular}& \begin{tabular}[c]{@{}l@{}}A review on LDA\end{tabular}  Mining software repositories using topic models & Unsupervised \\ [0.5ex]
        \hline
        \cite{thomas2014studying} & 2014 & 35 &SCP& Y & Y & N  & \begin{tabular}[c]{@{}l@{}}Explored Configurations without any explanation.\end{tabular}& \begin{tabular}[c]{@{}l@{}}Software Artifacts Analysis\end{tabular} & Unsupervised \\ [0.5ex]
        \hline
        \cite{chen2012explaining} &2012& 35 &MSR & Y & Y & N  & \begin{tabular}[c]{@{}l@{}}Choosing the optimal number of topics is difficult.\end{tabular}& \begin{tabular}[c]{@{}l@{}}Software Defects Prediction\end{tabular} & Unsupervised \\ [0.5ex]
        \hline
        \cite{thomas2014static} &2014& 31 &ESE& Y & Y & N  & \begin{tabular}[c]{@{}l@{}}Choosing right set of parameters is a difficult task.\end{tabular}& \begin{tabular}[c]{@{}l@{}}Software Testing\end{tabular} & Unsupervised \\ [0.5ex]
        \hline
        \cite{bajracharya2009mining} &2009 &29 & MSR& Y & Y & N  & \begin{tabular}[c]{@{}l@{}}Explored Configurations without any explanation \\and accepted to the fact their results were better \\because of the corpus they used.\end{tabular}& \begin{tabular}[c]{@{}l@{}}Software History \\ Comprehension\end{tabular} & Unsupervised \\ [0.5ex]
        \hline
        \cite{lohar2013improving} &2013& 27 &ESEC/FSE &Y & Y & Y  & \begin{tabular}[c]{@{}l@{}}Explored Configurations using LDA-GA.\end{tabular}& \begin{tabular}[c]{@{}l@{}}Traceability Link recovery\end{tabular} & Supervised \\ [0.5ex]
        \hline
        \cite{binkley2014understanding} &2014& 20 &ICPC& Y & Y & N  & \begin{tabular}[c]{@{}l@{}}Use heuristics to find right set of parameters.\end{tabular}& \begin{tabular}[c]{@{}l@{}}Source Code Comprehension\end{tabular} & Unsupervised \\ [0.5ex]
        \hline
        \cite{linares2013exploratory} &2013& 20 &MSR& Y & Y & N  & \begin{tabular}[c]{@{}l@{}}In Future, they planned to use LDA-GA.\end{tabular}& \begin{tabular}[c]{@{}l@{}}Stackoverflow Q\&A data \\ analysis\end{tabular} & Unsupervised \\ [0.5ex]
        \hline
        \cite{koltcov2014latent} & 2014 & 15 & WebSci& Y & Y & N  & \begin{tabular}[c]{@{}l@{}}Explored Configurations without any explanation.\end{tabular}& \begin{tabular}[c]{@{}l@{}}Social Software Engineering\end{tabular} & Unsupervised \\ [0.5ex]
        \hline
        \cite{grant2013using} & 2013 & 13 &SCP& Y & Y & N  & \begin{tabular}[c]{@{}l@{}}Their work focused on optimizing LDA’s topic \\count parameter.\end{tabular}& \begin{tabular}[c]{@{}l@{}}Source Code Comprehension\end{tabular} & Unsupervised \\ [0.5ex]
        \hline
        \cite{hindle2012relating} & 2012 & 13 &ICSM& Y & Y & N  & \begin{tabular}[c]{@{}l@{}}Explored Configurations without any explanation.\end{tabular}& \begin{tabular}[c]{@{}l@{}}Software Requirements \\ Analysis\end{tabular} & Unsupervised \\ [0.5ex]
        \hline
        \cite{fu2015automated} &2015& 6 & IST & Y & Y & N  & \begin{tabular}[c]{@{}l@{}}Explored Configurations without any explanation. \\Choosing right set of parameters is a difficult task.\end{tabular}& \begin{tabular}[c]{@{}l@{}}Software re-factoring\end{tabular} & Supervised \\ [0.5ex]
        \hline
        \cite{garousi2016citations} &2016& 5 &\begin{tabular}[c]{@{}c@{}}CS Review\end{tabular}& Y & Y & N  & \begin{tabular}[c]{@{}l@{}}Explored Configurations without any explanation.\end{tabular}& \begin{tabular}[c]{@{}l@{}}Bibliometrics and citations \\ analysis\end{tabular} & Unsupervised\\ [0.5ex]
        \hline
        \cite{le2014predicting} &2014& 5 & ISSRE& N & Y & N  & \begin{tabular}[c]{@{}l@{}}Explored Configurations without any explanation.\end{tabular}& \begin{tabular}[c]{@{}l@{}}Bug Localisation\end{tabular} & Semi-Supervised \\ [0.5ex]
        \hline
        \cite{nikolenko2015topic} & 2015 &3 &JIS& Y & Y & N  & \begin{tabular}[c]{@{}l@{}}They improvised LDA into ISLDA which gave \\stability across different runs.\end{tabular}& \begin{tabular}[c]{@{}l@{}}Social Software Engineering\end{tabular} & Unsupervised \\ [0.5ex]
        \hline
        \cite{sun2015msr4sm} &2015& 2 &IST& Y & Y & Y  & \begin{tabular}[c]{@{}l@{}}Explored Configurations using LDA-GA.\end{tabular}& \begin{tabular}[c]{@{}l@{}}Software Artifacts Analysis\end{tabular} & Supervised \\ [0.5ex]
        \hline
        \cite{chen2016topic} &2016& 0 &JSS& N & Y & N  & \begin{tabular}[c]{@{}l@{}}Explored Configurations without any explanation. \\Choosing right set of parameters is a difficult task.\end{tabular}& \begin{tabular}[c]{@{}l@{}}Software Defects Prediction\end{tabular} & Unsupervised\\ [0.5ex]
        \hline
\end{tabular}
\end{table*}

\begin{table*}[!t]
\renewcommand{\baselinestretch}{0.75}
\begin{center}
\footnotesize
\caption{LDA topic instability. Shows results from two runs. Column1 reports the percent
of words from a topic seen in its nearest match in the second run.}
\label{tbl:olap}
\begin{tabular}{r|l|l}
\% overlap with & &\\
closest topic in run2  & Topic name & top 9 words in topic\\\hline
100 &Xaml Binding & grid window bind valu wpf silverlight control xaml properti\\
100 & Ruby on Rails & rubi rail gem lib user end rvm app requir  \\ 
100 &MVC & http com java apach bean org springframework sun servlet\\
88 &Objective C & self cell nsstring iphon nil object anim alloc view\\
88 &Function Return Types & function const char void includ amp return int std\\
77 &Testing & chang tri work use problem like set code test\\
77 &Socket Communications & Main send socket connect run Android Activity thread start task process time\\
77 &Q and A & know need want way use question time like make\\
77 &OO Programming & instanc void type new method class return object public\\
77 &HTML Links & com http www url html site content page link\\
77 &Display & height jpg imag png src size img color width\\
66 &Windows/VS Tips & studio net dll web use mvc control asp visual\\
66 &Website Design & left span style text class div color css width\\
66 &Web Development & script jqueri function form input type ajax var javascript\\
66 &Java Programming & java void new privat return null int public string\\
66 &Date/Time Format & select day valu option year format month time date\\
66 &Database & column order group select join tabl row queri null\\
66 &Android User View & height content textview parent wrap android width view layout\\
66 &.NET Framework & net form text checkbox control click label asp button\\
55 &iOS App Development & applic user need iphon develop want app use like\\
55 &MySQL & connect sql databas tabl row queri data mysql server\\
55 &HTML Form & form login user usernam password page XML control action session\\
55 &Git Operations & folder upload open git use path read file directori\\
55 &Email Message & address form send field email messag error valid contact\\
55 &Eclipse & maven jar target build eclips version depend plugin project\\
44 &iOS App Development & applic user need iphon develop want app use like\\
44 &Regular Expressions & function array replac regex match echo php post string\\
44 &Media Player & object video play game player obj use data audio \\ 
44 &Float number Manipulation & doubl valu count rang number data length float point\\
44 &Compiling & librari compil lib includ usr command file error test\\
44 &.NET Framework & net form text checkbox control click label asp button\\
33 &Ruby Version Manager & end lib user rvm rubi rail app gem requir\\
33 &Scripting Language & def line self modul print import templat django python\\
33 &NodeJS & tag function node express parent use list like variabl\\
33 & C\# Programming & net web asp servic mvc use control applic wcf \\ 
33 & Python Programming & line python file print command script curl output run \\ 
33 &Android Debugging & com java lang debug method info android error androidruntim\\
22 & Visual Studio & window project file error librari compil visual build dll \\ 
22 & Information Systems & messag email log address contact send mail phone locat \\ 
22 &Flex4 Development & flash function new var list click sub event item
\end{tabular}
\end{center}
\end{table*}

\subsection{LDA is Widely Used}

We study LDA since this algorithm
is a widely-used technique in recent research papers appearing in prominent SE venues.
Tables~\ref{tab:venues}~\cite{sun2016exploring} and ~\ref{tbl:survey2} show top SE venues that published SE results and a sample of those papers respectively. For details on how Table~\ref{tbl:survey2} is generated, see Section \tion{solutions}.

As to {\em how} LDA is used, it is important to understand the distinction between supervised
and unsupervised data mining algorithms. In the general sense, most data mining is ``supervised'' when you have data samples associated with labels and then use machine learning tools to make predictions. For example,  in
the case of~\cite{lohar2013improving, sun2015msr4sm}, the authors used LDA as a feature extractor to build
feature vectors which, subsequently, were passed to a learner to predict for a target class. Note that such
supervised LDA processes can be fully automated and do not require human-in-the-loop insight to 
generate their final conclusions.

However, in case of,
``unsupervised learning'', the final conclusions are generated via a manual analysis and reflection over, e.g.,
the topics generated by LDA.  
Such cases represent~\cite{barua2014developers, hindle2012relating} who used a manual analysis of the topics generated from some SE tasks textual data by LDA as part of the reasoning within the
 text of that paper. It is possible to combine manual and automatic methods, please see the ``Semi-supervised''
 LDA paper of Le et al.~\cite{le2014predicting}.

However, as shown in our sample, one observation could be made that out of the 28 studies in Table \ref{tbl:survey2}, 23 of them make extensive
use of LDA for the purposes of {\em unsupervised exploration}. At some point these papers, browsed the LDA topics to guide their subsequent analysis. For example: 1) Barua et al.~\cite{barua2014developers} used the LDA topics to summarize the topics and trends in Stackoverflow, 2) Galvis et al.~\cite{galvis2013analysis} used LDA to gain an understanding of the nature of user statements in requirements documents, and many more.

As witnessed by the central columns of Table~\ref{tbl:survey2},
many prior papers~\cite{panichella2013effectively,lohar2013improving,sun2015msr4sm} have commented that the results of a topic modeling
analysis can be affected by tuning the control parameters of LDA.
Yet as reported in Section \tion{solutions},
a repeated pattern in the literature is that, despite these
stated concerns, researchers rarely take the next step to find ways to find better control tunings.

\subsection{Standard LDA Can Make Misleading Conclusions}
Standard practice in papers that use LDA is to present a table showing the top, say, 40
topics~\cite{barua2014developers}.
This section shows one example where, due to LDA instability,
the contents of such tables can only be described as  mostly illusionary.
Later in this paper (in Section \tion{unstable}) we show that this example is actually representative of
a general problem: changing the input ordering dramatically changes the topics
reported by LDA.

Barua et al.~\cite{barua2014developers}
analysed the topics and trends in Stackoverflow in which they reported the top 40 topics found at Stackoverflow. Table~\ref{tbl:olap} shows our attempt to reproduce their results.
Note that we could not produce a verbatim reproduction since they used the data of Stackoverflow dumped on 2012 while we used the current dump of 2016.
The analysis in Table~\ref{tbl:olap} is generated by running LDA twice with different
randomly generated input orderings (keeping all other parameter settings intact). After that, the topics generated from the first run where scored
according to the overlap of their words
from the second run.

We observed that while a very few topics appear verbatim in the two runs (see the top three lines of Table~\ref{tbl:olap}), but
most do not. In fact,
25 of the 40 topics in that figure show  instability (have an overlap of 55\% or less).

In Table~\ref{tbl:olap}, we only
examine the top $n=9$ words in each topic of run1 and run2. We selected \mbox{$n=9$} since,
later in this paper, we find that an analysis of \mbox{$n>9$} words per topics leads to near
 zero percent overlap of the topics generated in different runs (see Section \tion{unstable}).
That is, the instabilities of Table~\ref{tbl:olap} get even {\em worse} if we use {\em more} of the LDA output.

\subsection{LDA Stabilization Means Better Inference}

Inference with LDA can be assessed via topic similarities (as done in Table~\ref{tbl:olap}) and via the classification performance if the LDA topics are used as features to be fed into a classifier. As shown later in this paper, we can use LDADE to increase the similarities of the LDA
topics generated by LDA (see Section \tion{stable}).

As to the effects on classification accuracy, 
one way to score a classifier is via the $F_\beta$ score that combines
{\em precision} $p$ and {\em recall} $r$ as follows:
\begin{equation}\label{eq:f}
F_{\beta} = (1+\beta^2) \frac{pr}{p\beta^2 + r}
\end{equation}

We compared the analysis with $F_1$ ($\beta=1$) and $F_2$ ($\beta=2$) metrics seen in text mining classification using standard and stable topics generated by LDADE. The $F_2$ score is useful when reporting classification results
since it favors classifiers that do not waste the time on false positives. In Section \tion{rq3}, we report 
significant and large improvements in both the evaluation metrics. That is, LDADE not only improves topic stability, but also the efficacy
of the inference that subsequently uses those topics.

\section{Related Work}
\label{sect:related}

\subsection{Topic Modeling}\label{sect:tm}

LDA is a generative statistical model that allows
sets of observations to be explained by unobserved groups that explain why some
parts of the data are similar. It learns the various distributions (the set of
topics, their associated word probabilities, the topic of each word, and the
particular topic mixture of each document).
What makes topic modeling interesting is that these algorithms scale to very
large text corpuses.  For example, in this paper, we apply LDA to whole of Stackoverflow,
as well as to two other large text corpuses in SE.
\begin{table}[!t]
\scriptsize
\begin{center}
\caption{Example (from~\cite{barua2014developers}) 
of generating topics from Stackoverflow. For each topic, we show just the five most heavily
  weighted words.}
 \label{tbl:lda}
\begin{tabular}{c|c|c|c|c}
 
        \begin{tabular}[c]{@{}c@{}}Topic: String\\ Manipulation\end{tabular}    &\begin{tabular}[c]{@{}c@{}}Topic:\\ Function\end{tabular}    &\begin{tabular}[c]{@{}c@{}} Topic: OO \\Programming\end{tabular} &\begin{tabular}[c]{@{}c@{}}Topic: UI \\Development\end{tabular}&\begin{tabular}[c]{@{}c@{}}Topic: File \\Operation\end{tabular} \\\hline
string&function& class& control&file \\
charact&paramet& method& view&directori\\
encod&pass& object& event&path\\
format&return& call& button&folder\\
convert&argument& interfac& click&creat\\\hline
\end{tabular}
\end{center}
\end{table}

Table~\ref{tbl:lda} illustrates topic generation from Stackoverflow.
To find these topics, LDA explores two probability distributions: 
\bi
    \item $\alpha=P(k|d)$, probability of topic $k$ in  document $d$;
    \item $\beta=P(w|k)$, probability of word $w$ in topic $k$.
\ei
\noindent
Initially, $\alpha$ and $\beta$ may be set randomly as follows:
each word in a document was generated by first randomly picking a topic (from
the document’s distribution of topics) and then randomly picking a word (from
the topic’s distribution of words). Successive iterations of the algorithm 
count the implications of prior sampling which, in turn,  incrementally updates $\alpha$ and $\beta$.

Binkley et al.~\cite{binkley2014understanding} performed an extensive study and found that 
apart from $\alpha$ and $\beta$, the other parameters that define LDA
are: 
\bi
    \item $k$ = number of topics
    \item $b$ = number of burn-in iterations;
    \item $si$ = the sampling interval.
\ei
\noindent
Binkley et al.'s study of the LDA settings was a mostly manual process
guided by their considerable background knowledge and expertise and program
comprehension.
In the field of program comprehension, the Binkley article
is the state of the art in applications of LDA to software engineering.

To that work, this paper adds a few extra conclusions.
Firstly, we explore LDA in fields other than program comprehension.
Secondly, we ask the question ``what if the analysts lacks extensive background knowledge
of the domain?''. In that circumstance, some automatic method is needed to support
an informed selection of the LDA parameters.

 

\subsection{About Order Effects}

\noindent
This paper uses tuning to fix ``order effects'' in topic modeling. Langley~\cite{gennari1989models} defines such effects as follows:
\begin{quote}
{\em A learner $L$ exhibits an order effect on a training set  $T$ if there exist
two or more orders of $T$ for which $L$ produces different knowledge structures.}
\end{quote}
Many learners exhibit order effects, e.g., certain incremental clustering algorithms generate different
clusters, depending on the order with which they explore the data~\cite{gennari1989models}.
Hence, some algorithms survey the space of possible models across numerous
random divisions of the data (e.g., Random Forests~\cite{breiman2001random}).

From the description offered above in Section \ref{sect:tm},
we can see how topic modeling might be susceptible to order effects and how such order
effects might be tamed: 1) In the above description, $k$, $\alpha$ and $\beta$ are initialized at random
then updated via an incremental re-sampling process. Such incremental updates are prone to order effects, and 2) other way is to initialize, $k$, $\alpha$ and $\beta$ to some
  useful value. As shown in Section \tion{diff},
the key to  applying this technique is that different data sets will require different
  initializations, i.e., the tuning process will have to be repeated for each new data set.

\subsection{Tuning: Important and (Mostly) Ignored}
\label{sect:tune}

The impact of tuning is well understood in the theoretical machine learning literature~\cite{bergstra2012random}.  When we tune a
data miner, what we are really doing is changing how a learner applies its
heuristics. This means tuned data miners use different heuristics, which means
they ignore different possible models, which means they return different models,
i.e., \textit{how} we learn changes \textit{what} we learn.

Yet issues relating to
tuning are poorly addressed in the software analytics literature.  Fu et al.~\cite{fu2016tuning} surveyed hundreds of recent SE papers in the area
of software defect prediction from static code attributes. They found that most SE
  authors do not take steps to explore tunings (rare exception:~\cite{tantithamthavorn2016icse}). For example, Elish et
  al~\cite{elish2008predicting} compared support vector machines to other data
  miners for the purposes of defect prediction. That paper tested different
  ``off-the-shelf'' data miners on the same data set, without adjusting the
  parameters of each individual learner. Similar comparisons of data miners in SE,
with no or minimal pre-tuning study, can be found in the work on Lessmann et al.~\cite{lessmann2008benchmarking}
and, most recently, in Yang et al~\cite{yang2016effort}.  

In summary, all our literature reviews of the general  (non-LDA) software analytics literature
show that
the importance of tuning is often mentioned, but never directly addressed.

\subsection{LDA,  Instability and Tuning}
\label{sect:solutions}
Within the LDA literature, some researchers
have explored LDA instability.
We searched scholar.google.com for papers published before August 2016, for the conjunction of ``lda'' and ``topics'' or ``stable'' or
``unstable'' or ``coherence''. Since 2012, there are  189 such papers, 57
of which are related to software engineering results. Table \ref{tbl:survey2} gives a broad discussion on these papers. In short, of those papers:
\bi
\item 28/57
mention instability in LDA. 
\item Of those 28, despite mentioning stability problems,
  10 papers still used LDA's ``off-the-shelf'' parameters;
  \item The  other 28-10=18 papers used some combination of manual adjustment or some
under-explained limited exploration of tunings based on ``engineering judgment''
(i.e., some settings guided by the insights of the researchers).
\item
Only 4 of the authors acknowledge that tuning might have a large impact
on the results.
\ei
Apart from tuning, there are several other workarounds explored in the literature
in order to handle LDA instability. Overall, there was little systematic exploration of tuning and LDA in the SE literature.
Instead, researchers relied on other methods that are less suited to automatic reproduction of prior results.

In the literature, researchers~\cite{maskeri2008mining, martin2015app, guzman2014users}
    manually accessed the topics and then used for further experiments. Some
    made use of Amazon Mechanical Turk to create gold-standard coherence
    judgements~\cite{lau2014machine}. All these solutions are related to results
    stability rather than model stability.
    Note that this workaround takes extensive manual effort and time.

    Another approach to tame LDA instability
    is to incorporate
    user knowledge into the corpus. For example,
    SC-LDA~\cite{yang2015improving} can
    handle different kinds of knowledge such as word correlation,
    document correlation, document label and so on. Using such user
    knowledge, while certainly valuable, is somewhat subjective.
    Hence, for reasons of reproducibility, we prefer fully
    automated methods.

Some researchers 
used
genetic
algorithms to learn better settings for LDA~\cite{panichella2013effectively,lohar2013improving,sun2015msr4sm}.
Genetic algorithms are 
themselves a stochastic search process. That is, the changes to 
input orderings explored in this paper would introduce further conclusion instability
from the genetic algorithms.
In principle, that instability could be removed via extra  runs of genetic algorithms 
over multiple sub-samples of that, where the GA goals are augmented to include
``similar topics should be found in different runs''.
That said:
\bi
\item None of the prior work using GAs to improve LDA have applied those sub-sampling stability test;
\item If done naively, adding further goals and data sub-sampling to a GA runs the risk
  of dramatically increasing the runtimes.
One reason to prefer LDADE is that it terminates very quickly.
\ei
Finally, other researchers explore
some limited manual parameter tuning for LDA
(e.g., experiment with one parameter: cluster size)~\cite{galvis2013analysis, tian2009using}
to achieve higher stability by just increasing the number of cluster size.
Note that the automatic tuning methods explored by this paper can
explore multiple parameters. Further, our analysis is repeatable.

\section{Methods}
\label{sect:evaluation}
This section describes our evaluation methods for measuring instability as well as the optimization
methods used to reduce that instability. All our code scripts, methods and results can be found online\footnote{https://github.com/ai-se/Pits\_lda/}.

\subsection{Data Sets}
To answer our research questions, and to enable reproducibility of our results,
we use three open source datasets summarized in Table~\ref{tbl:dataset} and described
below. These 3 datasets are unrelated which solve different SE tasks. We wanted to make sure our LDADE is useful for these 3 tasks. This puts emphasis on the importance of stability in LDA. 

\begin{table}[!htbp]
\scriptsize
\begin{center}
\caption{Statistics on our datasets. PitsA, PitsB, etc refer to the issues
from six different NASA projects.}
\label{tbl:dataset}
\begin{tabular}{c|c|c}
  \begin{tabular}{cc@{}c@{}}
    \multicolumn{2}{c}{\textbf{~}}
  \end{tabular}
  & \multicolumn{2}{c}{\textbf  Size} \\
    \cline{2-3}
         & \textbf{Before} & \textbf{After}\\ 
  \textbf{Data set}      & \textbf{Preprocessing} & \textbf{Preprocessing}\\ 
        \hline
        PitsA & 1.2 MB & 292 KB \\ 
        \hline
        PitsB & 704 KB & 188 KB \\
        \hline
        PitsC & 143 KB & 37 KB \\ 
        \hline
        PitsD & 107 KB & 26 KB \\ 
        \hline
        PitsE & 650 KB & 216 KB \\
        \hline
        PitsF & 549 KB & 217 KB \\ 
        \hline
        Citemap & 8.6 MB & 3.7 MB \\ 
        \hline
        Stackoverflow & 7 GB & 589 MB \\ 
\end{tabular}
\end{center}
\end{table}

\begin{figure*}[!t]
  \begin{center}
    \includegraphics[width=13cm]{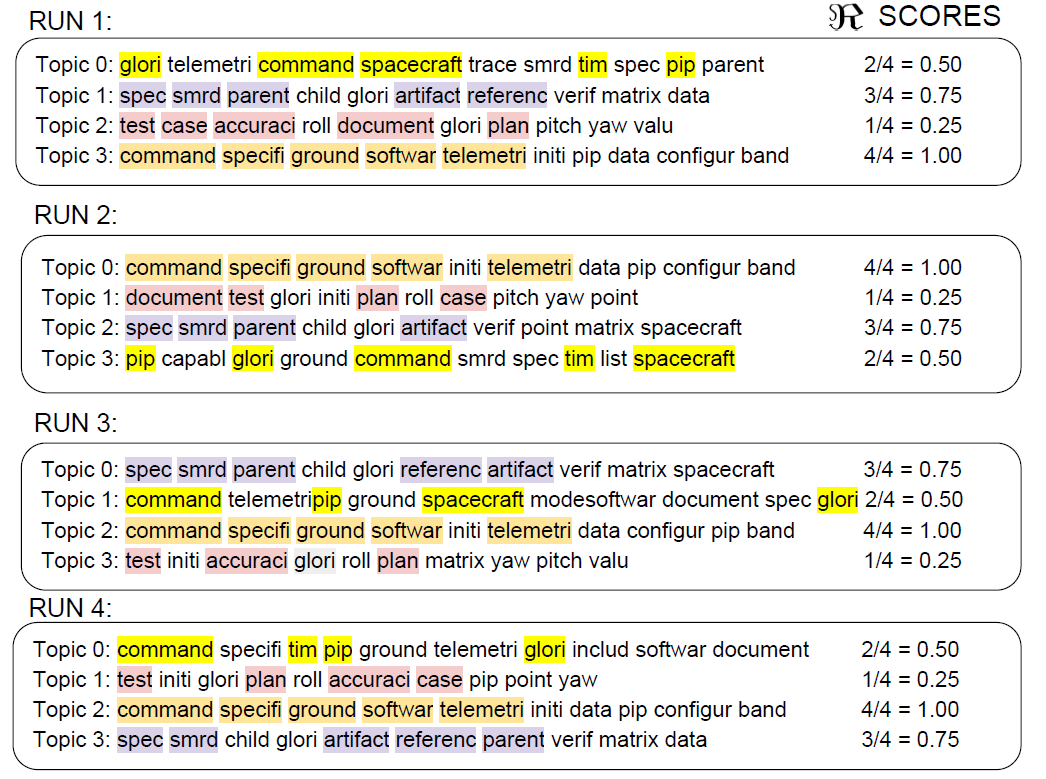}
    \end{center}
  \caption{Example of topics overlap off size $n=5$ across multiple runs.}
  \label{fig:jaccard}
\end{figure*}

\textbf{PITS} is a text mining data set generated from NASA software project
and issue tracking system (PITS) reports~\cite{menzies2008improving, menzies2008automated}. This text discusses
bugs and changes found in big reports and  review patches.
Such issues are used
to manage quality assurance, to support communication
between developers. Topic modeling in PITS can be used
to identify the top topics which can
identify each severity separately. The dataset can be downloaded from the
PROMISE
repository~\cite{promiserepo}. Note that, this data comes from six different
NASA projects, which we label as PitsA, PitsB, etc.
    
 \textbf{Stackoverflow} is the flagship site of the Stack Exchange Network which
 features questions and answers on a wide range of topics in computer
 programming. There has been various studies done to find good topics on Stackoverflow for SE~\cite{barua2014developers,linares2013exploratory, allamanis2013and,rosen2016mobile}.
Topic modeling on Stackoverflow is useful for finding patterns in programmer knowledge.
 This data can be downloaded online\footnote{http://tiny.cc/SOProcess}. 
    
  \textbf{Citemap} contains titles and abstracts of 15121 papers from a
 database of 11 senior software engineering conferences from 1992-2016. Most of this data was
 obtained in the form of an SQL dump from the work of Vasilescu et
 al.~\cite{vasilescu2013historical} and some are collected by Mathew et al~\cite{mathew2017trends}. People have studied healthiness of software engineering conferences~\cite{vasilescu2014healthy}. This dataset is available online\footnote{https://github.com/ai-se/Pits\_lda/blob/master/dataset/citemap.csv}.

  For this study, all  datasets were preprocessed using the usual text mining filters~\cite{feldman2006j}:
\bi
\item
  Stop words removal using NLTK toolkit\footnote{http://www.nltk.org/book/ch02.html}~\cite{bird2006nltk} : ignore very common short words such as  ``and'' or ``the''.
\item
  Porter's stemming filter~\cite{Porter1980}: delete uninformative word endings, e.g., after performing stemming, all the following words would be rewritten
  to ``connect'': ``connection'', ``connections'',
``connective'',          
``connected'',
  ``connecting''.
\item
  Tf-idf feature selection: focus on the 5\% of words that occur frequently,
  but only in small numbers of documents. If a word occurs $w$ times
  and is found in $d$ documents  and there
  are $W,\ D$ total number of words and documents respectively, then tf-idf is scored
  as follows:
  \[
  \mathit{tfidf}(w,d)=   \frac{w}{W} *\log{\frac{D}{d}}\]
  \ei

  Table~\ref{tbl:dataset} shows the sizes of our data before and after pre-processing.
  These datasets are of different sizes and so are processed using different tools: 1) PITS and Citemap is small enough to process on a single (four core) desktop machine
  using Scikit-Learn~\cite{pedregosa2011scikit} and Python, and 2) Stackoverflow is so large (7GB) that its  processing requires extra hardware support.
 This study used Spark and Mllib on a cluster of 45 nodes to
 reduce the runtime.

\subsection{Similarity Scoring}

To evaluate topics coherence in LDA, there is a direct approach, by asking people about topics, and an indirect approach by evaluating \textit{pointwise mutual information (PMI)}~\cite{lau2014machine, o2015analysis} between the topic words. We could not use any of these criteria, as it requires experts to have domain knowledge. \textit{Perplexity} is  the inverse of the geometric mean per-word likelihood. The smaller the perplexity, the better (less uniform) is the LDA model. The usual trend is that as the value of perplexity drops, the number of topics should grow~\cite{koltcov2014latent}. Researchers caution that the value of perplexity does not remain constant with different topic and corpus sizes~\cite{ zhao2015heuristic}. Perplexity depend on its code implementation and the type of datasets used. Since we are using different implementations of LDA across different platforms on various datasets, we are not using perplexity as evaluation measure.

There is well known measure, called \textit{Jaccard Similarity}~\cite{o2015analysis, galvis2013analysis}, for measuring similarity. But we modified the measure to do a cross-run similarity of topics. For this work, we assess topic model stability via the {\em median number overlaps of size $n$ words $\mathit{(size\_of\_topic)}$}, which is denoted as $\Re_n$.

For this measurement, we first determine the maximum size of topics we will study. For that purpose,
we will study the case of $n \le 9$ (we use 9 as our maximum size since the cognitive
science literature tells us that $7\pm 2$ is a useful upper size for artifacts to be browsed by humans~\cite{miller1956magical}).

Next, for $1 \le n \le 9$, we will calculate the median size of the overlap,
computed as follows:
\bi
\item Let one {\em run} of our rig shuffle the order of the training data, then build topic models using the data;
  \item $m$ runs of our rig execute $m$ copies of one run, each time using a different random number seed,
\item Topics are said to be stable,
when there are $x$ occurrences of  $n$ terms appearing in all the topics seen in the $m$ runs.
\ei

For example, consider the topics shown in Figure~\ref{fig:jaccard}. These are generated via four {\em runs} of our system. In this hypothetical example, we will assume that the runs of
 Figure~\ref{fig:jaccard} were generated by an LDA suffering from topic instability.
For $n=5$, we note that Topic~0 of run1 scores $\frac{2}{4}=0.5$ since it shares 5 words with topics in only two out of four runs.
Repeating that calculation for the other run1 topics shows that:
\bi
\item Topic~1 of run1 scores $\frac{3}{4}=0.75$;
\item Topic~2 or run1 scores $\frac{1}{4}=0.25$;
\item Topic~3 of run1 scores $\frac{4}{4}=1$.
  \ei
  From this information, we can calculate
  $\Re_5$  (the
  {\em median number overlaps of size $n=5$ words}) as:
  \[
   \mathit{median}(0.5, 0.75, 0.25, 1) =0.625\]

\begin{figure}[!h]
  \includegraphics[width=\linewidth]{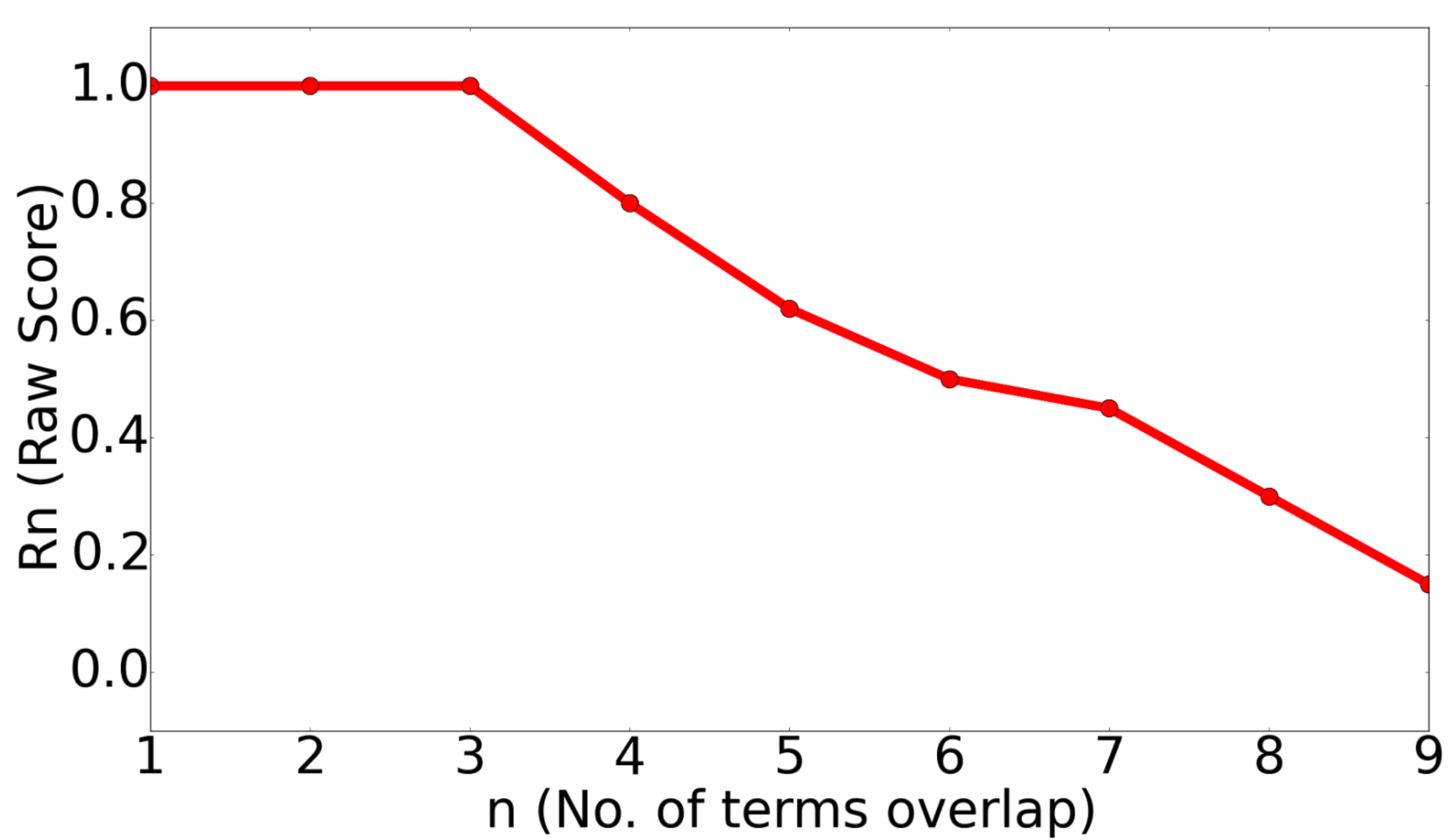}
  \caption{$\Re_n$ scores of 
  Figure~\ref{fig:jaccard} for $1 \le n \le 9$}
  \label{fig:alln}
\end{figure}

  Figure~\ref{fig:alln}
  shows the $\Re_n$ scores of 
  Figure~\ref{fig:jaccard} for $1 \le n \le 9$.  From this figure, we can see LDA topic instability
  since
  any report of the contents of a topic that uses more than three words per topic would be unreliable.

 For the following analysis,
we distinguish between the \textbf{\textit{Raw  score}} and the \textbf{\textit{Delta  score}}:
 \bi
\item The two \textbf{\textit{Raw  scores}} are the $\Re_n$ median similarity scores seen {\em before} and {\em after} tuning LDA;
\item The \textbf{\textit{Delta score}} is the difference between the two
  \textbf{\textit{Raw scores}} (after tuning - before tuning).  \ei 
  The pseudocode for these calculations
  is shown in \fig{pseudo_un} (Algorithm 1) with the default set of parameters. In the following
  description, superscript numbers denote lines in the pseudocode. The data ordering is
  shuffled every time LDA is ran$^{7}$. Data is in the form of $\mathit{tfidf}$
  scores of each word per document. Shuffling is done in order to induce maximum
  variance among the ordering of data with  different runs of LDA. Function $\mathit{lda}^{8}$ returns $k$ topics. Topics$^{8}$ are a list of lists which
  contains topics from all the different runs. A stability score is evaluated on
  every 10 runs (Fixed) of LDA, and this process is continued 10 (Fixed) times to avoid any sampling bias. At the end, the median
  score is selected as the untuned raw score ($\Re_n$ ) $^{4-12}$. Hence, the runtimes comes from $10$ evaluations of untuned experiment.

    

\begin{figure}[!h]
\begin{lstlisting}[mathescape,linewidth=8.2cm,frame=r,numbers=right]
  # Algorithm 1
  def LDASCORE($n$, $k$, $\alpha$, $\beta$, $Data$):
    $Score$ = emptySet
    for j = 0 to 10 do
        $Topics$ = emptySet
        for i = 0 to 10 do
            data = shuffle($Data$)
            Topics.add(lda($k,\ \alpha,\ \beta,\ data$))
        end for
        $Score$.add($Overlap$(Topics, $n$, $k$))
    end for
    $Raw\_Score $ = median(Score)
    return $Raw\_Score $
\end{lstlisting}
\caption{Pseudocode for untuned LDA with Default Parameters}
\label{fig:pseudo_un}  
\end{figure}

\subsection{Tuning Topic Modeling with LDADE}
\label{sect:tuning}
LDADE is a combination of topic modeling (with LDA) and an optimizer (differential evolution, or DE) that adjusts
the parameters of LDA in order to optimize (i.e., maximize) similarity scores.

We choose to use DE after a literature search on search-based SE methods.
The literature mentions many optimizers: simulated
annealing~\cite{feather2002converging, menzies2007business}; various genetic
algorithms~\cite{goldberg1979complexity} augmented by techniques such as
DE (differential evolution~\cite{storn1997differential}), tabu search and scatter
search~\cite{glover1986general, beausoleil2006moss, molina2007sspmo,nebro2008abyss}; particle swarm optimization~\cite{pan2008particle}; numerous
decomposition approaches that use heuristics to decompose the total space into
small problems, then apply a response surface methods~\cite{krall2015gale, zuluaga2013active}.
Of these, we use DE for two reasons. Firstly, it has been proven useful in prior SE tuning
studies~\cite{fu2016tuning}. Secondly, our reading of the current literature is
that there are many advocates for differential evolution like Vesterstrom et al.~\cite{vesterstrom2004comparative} showed DE to be
competitive with particle swarm optimization and other GAs. 

LDADE  adjusts the parameters of
Table~\ref{tb:tuned}. Most of these parameters were explained above. Apart from them, there are 2 different kinds of LDA implementations as well and they are: 1) VEM is the deterministic {\em variational EM} method that computes $\alpha$ and $\beta$ via
  expectation maximization~\cite{minka2002expectation}, and 2) Gibbs sampling~\cite{wei2006lda, griffiths2004finding} is a Markov Chain Monte Carlo algorithm, which is an approximate stochastic process for computing and updating $\alpha$ and $\beta$.
  Topic modeling researchers in SE have argued that Gibbs leads to stabler models~\cite{layman16a,layman2016topic} (a claim which we test, below).

We manually run these other inference techniques according to different implementations. We need to make sure that these instabilities do not hold for just 1 inference technique, or 1 implementation.

\begin{table}[!htbp]
\begin{center}
\scriptsize
\caption{List of parameters tuned by this paper}
\label{tb:tuned}
\begin{tabular}{|c|c|c|p{3.5cm}|}
        \hline 
        \textbf{Parameters} & \textbf{Defaults} & \textbf{Tuning Range} & \textbf{Description}\\
        \hline
        $k$ & 10 & [10,100] & Number of topics or cluster size \\ 
        \hline
       $\alpha$ & None & [0,1] & Prior of document topic distribution. This is called alpha \\ 
        \hline
        $\beta$ & None & [0,1] & Prior of topic word distribution. This is called  beta \\

        \hline
\end{tabular}
\end{center}
\end{table}

\fig{pseudo_DE} (Algorithm 2) shows the pseudocode of LDADE. DE evolves from the \textit{NewGen} of
candidates from a current $Pop$. Each candidate solution in the $Pop^{10}$ is a set of  parameters (Tunings). The values of this set is selected randomly from Table~\ref{tb:tuned} in the $\mathit{InitializePopulation}^{10}$ function. $Pop$ variable is now a matrix of $10\times3$ since $np=10$.

\begin{figure}[!h]
\scriptsize
\begin{lstlisting}[mathescape,linewidth=8.2cm,frame=r,numbers=right]
  # Algorithm 2
  def LDADE($np=10$,  # size of frontier
            $f=0.7$,  # differential weight
            $cr=0.3$, # crossover probability
            $iter=3$, # number of generations
            $n$,      # words per topic
            $Data$,   # tf*idf scores
            Goal $\in$ Maximizing $\Re_n$ (Raw) score)
    $Cur\_Gen$ = emptySet
    $Pop$ = $InitializePopulation(np)$ # A matrix of 10 by 3
    for $i$ = 0 to $np-1$ do
        temp = $ldascore$($n$, $Pop$[i], $Data$)
        $Cur\_Gen$.add([$Pop$[i], temp])
    end for
    for $i$ = 0 to $iter$ do
        $NewGen$ = emptySet
        for $j$ = 0 to $np-1$ do
            $S_i$ = Extrapolate($Pop$[j], $Pop$, n, cr, f, np)
            temp = ldascore(n, $S_i$, Data)
            if temp $\geq$ $Cur\_Gen$[j][1] then
              $NewGen$.add([$S_i$, temp])
            else
              $NewGen$.add([$Cur\_Gen$[j][0], $Cur\_Gen$[j][1]])
            end if
        end for
        $Cur\_Gen$ = $NewGen$
    end for
    $Raw\_Score,\ best\_set$ = GetBestSolution($Cur\_Gen$)
    return $Raw\_Score,\ best\_set$
  
  def Extrapolate($old,\ pop,\ cr,\ f,\ np$)
        $a,\ b,\ c$ = $threeOthers$(pop) # select any 3 items
        $newf$ = emptySet
        for $i$ = 0 to $np-1$ do
            if $cr \leq$ random() then
                $newf$.add($old[i]$)
            else
                $newf$.add(trim(i, ($a$[i]+$f\ast$($b$[i] $-$ $c$[i]))))
            end if
        end for
        return $newf$ 
\end{lstlisting}
\caption{Pseudocode for DE with a constant number of iterations}
\label{fig:pseudo_DE}  
\end{figure}

$\mathit{Cur\_gen}^{9}$ and $\mathit{NewGen}^{16}$ variables are the list of Tunings, and $\Re_n$ score 
which comes similarly from Algorithm 1$^{4-12}$ (\fig{pseudo_un}). But for LDADE, everytime Algorithm 1 uses different values of parameters found by DE. The runtimes comes from 1 DE run which does about $\mathit{iter} * np$ evaluations of tuned experiment. DE is driven by a goal (called quality/fitness funtion), which in this case is maximizing the $\Re_n$ (raw) score calculated using  $\mathit{ldascore}^{19}$ function.

The main loop of DE$^{15}$ runs over the \textit{Pop}, replacing old items with new Candidates (if new candidate is better).
DE generates \textit{new Candidates} via 
$\mathit{Extrapolate}^{18}$ function between current solutions of the $\mathit{Cur\_gen}$ variable. Three solutions $a$, $b$ and $c\ ^{32}$ are
selected at random from the $\mathit{pop}$. Each of these solution is nothing but a set of ($k$, $\alpha$ and $\beta$). For each tuning parameter i$^{34}$, at some crossover probability (\textit{cr} $^{35}$), we
replace the $\mathit{old}$ tuning with $\mathit{newf}$. We mutate a solution with the equation $y_i = a_i + f \times (b_i - c_i)$ where $f$ is a
parameter controlling crossover. The trim function$^{38}$ limits the new value
to the legal range min..max of that parameter.

The loop invariant of DE is that, after the zero-th iteration$^{15}$, the \textit{Pop}
contains examples that are better than at least one other candidate$^{20}$.
As the looping progresses, the \textit{Pop} is full of increasingly more valuable solutions
which, in turn, also improve the candidates, which are Extrapolated from the Population. LDADE finds the optimal configuration and the $\Re_n$ (Raw) score$^{28}$ using $GetBestSolution$ function for a particular dataset to be used with LDA for further SE task. One may argue why we have been encoding $\Re_n$ score in the solution. This is shown in pseudocode just to keep track of the values but it is not used to drive our quality/fitness ($ldascore$) function.

\begin{table}[!htbp]
\begin{center}
\scriptsize
\caption{Overview of Algorithm LDADE}
\label{tb:algo2}
\begin{tabular}{|c@{~}|p{4.7cm}|}
        \hline 
        \textbf{Keywords} & \textbf{Description}\\
        \hline

        \begin{tabular}[c]{@{}c@{}}Differential weight $(f=0.7)$\end{tabular} & Extent of mutation to be performed on the candidates\\
        \hline
        \begin{tabular}[c]{@{}c@{}}Crossover probability $(cr=0.3)$\end{tabular} & Representation of the survival of the candidate in the next generation\\
        \hline
        \begin{tabular}[c]{@{}c@{}}Population Size $(np=10)$\end{tabular} &  Frontier size in a generation \\
        \hline
        \begin{tabular}[c]{@{}c@{}}No. of Generations $(iter=3)$\end{tabular} & How many generations need to be performed\\
        \hline
        \begin{tabular}[c]{@{}c@{}}Fitness Function $(ldascore)$\end{tabular} & Driving factor of DE\\
        \hline
        \begin{tabular}[c]{@{}c@{}}GetBestSolution Function\end{tabular} & Returns optimal configuration and it's corresponding $\Re_n$ score \\
        \hline
        \begin{tabular}[c]{@{}c@{}}Extrapolate Function\end{tabular} & Selects between the current candidate and next candidate\\
        \hline
        \begin{tabular}[c]{@{}c@{}}Output\end{tabular} & Optimal Configurations to use with LDA for further SE Task\\
        \hline
\end{tabular}
\end{center}
\end{table}

Table~\ref{tb:algo2} provides an overview of LDADE algorithm. Also, note that DEs have been
applied before for parameter tuning (e.g., see~\cite{omran2005differential,chiha2012tuning, fu2016tuning} ) but this is the first time they have been
applied to tune LDA to increase stability.

\section{Results}\label{sect:results}

In this section,
 any result from the smaller data sets (Pits and Citemap) come
from Python implementation based on Scikit-Learn running on a 4 GB ram machine (Linux).
Also,
  any results from the larger data (Stackoverflow) comes from a Scala implementation
  based on Mllib~\cite{meng2016mllib} running on a 45 node Spark system (8 cores per node).
  
  Note that, for the RQ3, there are some intricate details with classification results. After tuning (Goal is still to maximize the $\Re_n$ score) and finding the optimal ``k'', we trained a Linear Kernel SVM classifier using document topic distributions as features just like used by Blei et al~\cite{blei2003latent}.

\subsection{\textbf{RQ1: Are the default settings of LDA incorrect?}}\label{sect:unstable}

This research question checks the core premise of this work, that changes
in the order of training data dramatically affects the topics learned via LDA.
Note that if this is {\em not true}, then there would be no value added to this paper. 

  \begin{table*}[!t]
\renewcommand{\baselinestretch}{0.75}
\begin{center}
\footnotesize
\caption{LDADE finds 27 stable topics which were unstable with the use of LDA shown in Table~\ref{tbl:olap}. Column1 reports the percent
of words from a topic seen in its nearest match in the second run.}
\label{tbl:olap_stable}
\begin{tabular}{r|l|l}
\% overlap with & &\\
closest topic in run2  & Topic name & top 9 words in topic\\\hline
100 &C Programming & function return const char void includ amp int std\\
100 &Visual Studio & studio net visual dll web use window control asp\\
100 &Xaml Binding & grid window bind xaml valu data silverlight control properti\\
100 &HTML Form & form page user usernam password XML action session control login\\
100 &Objective C & alloc nsock nsstring view iphon cell nil object anim\\
88 &iOS Software & applic user need iphon develop want app use like\\
88 &Testing & code suite chang work use problem software set test\\
88 &Date/Time Format & month time date select day valu option year format\\
88 &Socket Communications & network protocol send socket connect run address thread start \\
88 & Ruby on Rails & gem rail lib user rubi end mvc app requir  \\
88 &.NET Framework & asp net form text checkbox control click label button\\
88 &OO Programming & class instanc void public type interface method return object\\
77 &Android User View & android view layout height content textview parent wrap width\\
77 &Java Envrionment & http java apach bean org spring framework sun servlet \\
77 &Display & png src size height jpg imag img color width\\
77 & CSS & color css width left span style text class div\\
77 &Email Message & messag error address form email send field  valid contact\\
77 &Web Development & ajax var javascript script jqueri function form input type\\
77 &Git Operations & git use path read folder directori upload open file \\
66 &Java Programming & java privat return null void new  int public string\\
66 &MySQL & tabl row queri data mysql connect sql databas server\\
66 &Regular Expressions & regex match echo php post function array replac string\\
66 &Eclipse & eclips version depend plugin maven jar target build  project\\
66 &Database & group join tabl row column order select  queri null\\
66 &iOS App Development & app iphon applic user need  develop want ios like\\
55 &HTML Links & url html com http www content page site  link\\
55 & Compiling & compil lib includ usr librari  command file error test\\
\end{tabular}
\end{center}
\end{table*}
 
\begin{figure}[!b]
  \begin{center}
    \includegraphics[width=\linewidth]{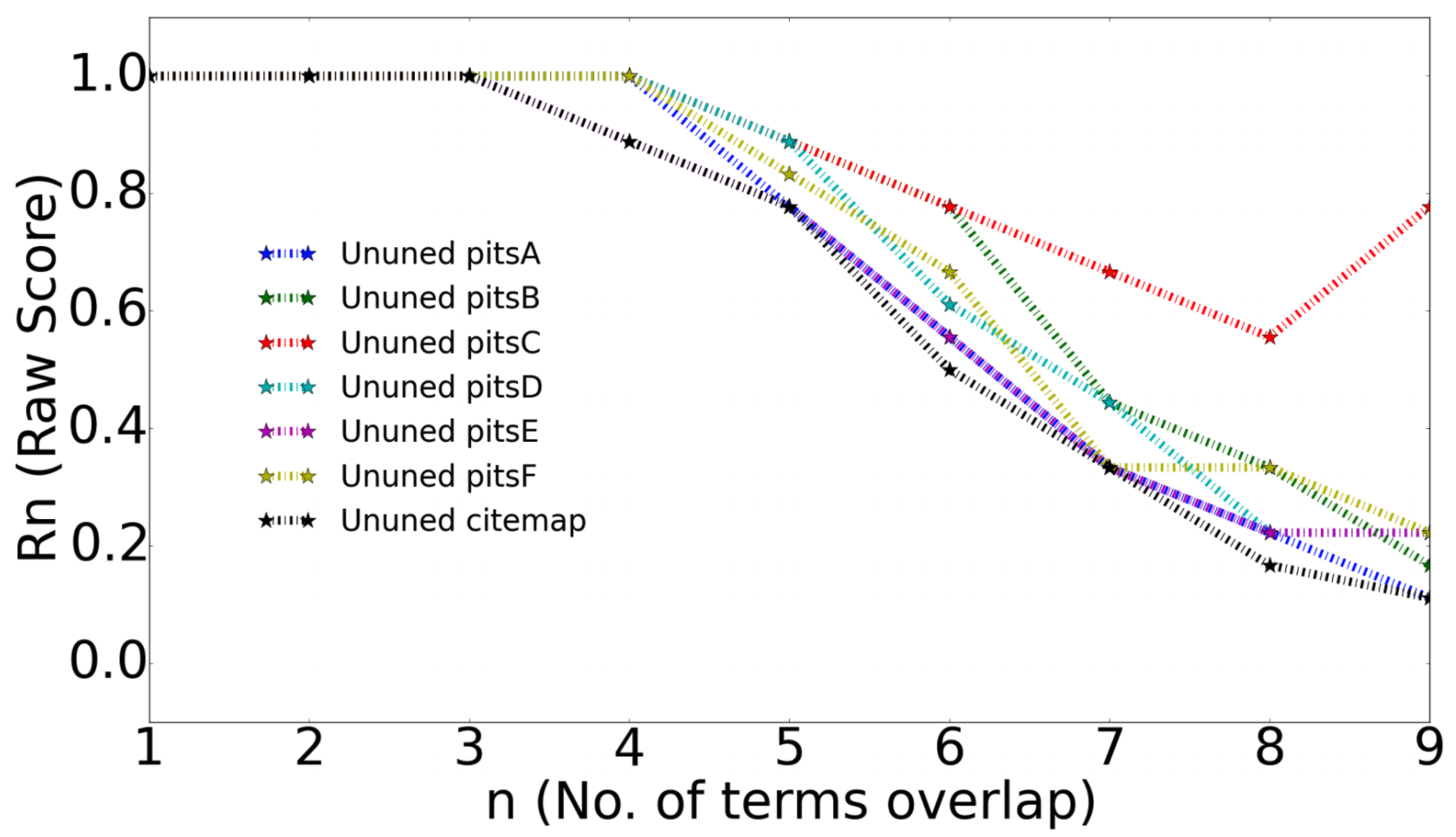}
    \end{center}
  \caption{{\em Before} tuning: uses LDA's default parameters}\label{fig:delta11}  
\end{figure}

Figure~\ref{fig:delta11}   plots $n$ vs $\Re_n$ for untuned  LDA.
Note that the  stability collapses the most after $n=5$ words. This means
  that any report of LDA topics that uses more than five words per topic will
  be changed, just by changing the order of the inputs. This is a significant result
  since the standard advice in the LDA papers~\cite{panichella2013effectively, lukins2010bug}
  is to report the top 10 words per topic. As shown in Figure~\ref{fig:delta}a, it would
  be rare that any such 10 word topic would be found across multiple runs.
 \begin{lesson}
  Using the default settings of LDA for these SE data can lead to systematic errors due to topic
  modeling instability. 
\end{lesson}

\subsection{\textbf{RQ2: Does LDADE improve the stability scores?}}\label{sect:stable}

\begin{figure}[!t]
        \begin{center}
        \includegraphics[width=0.9\linewidth]{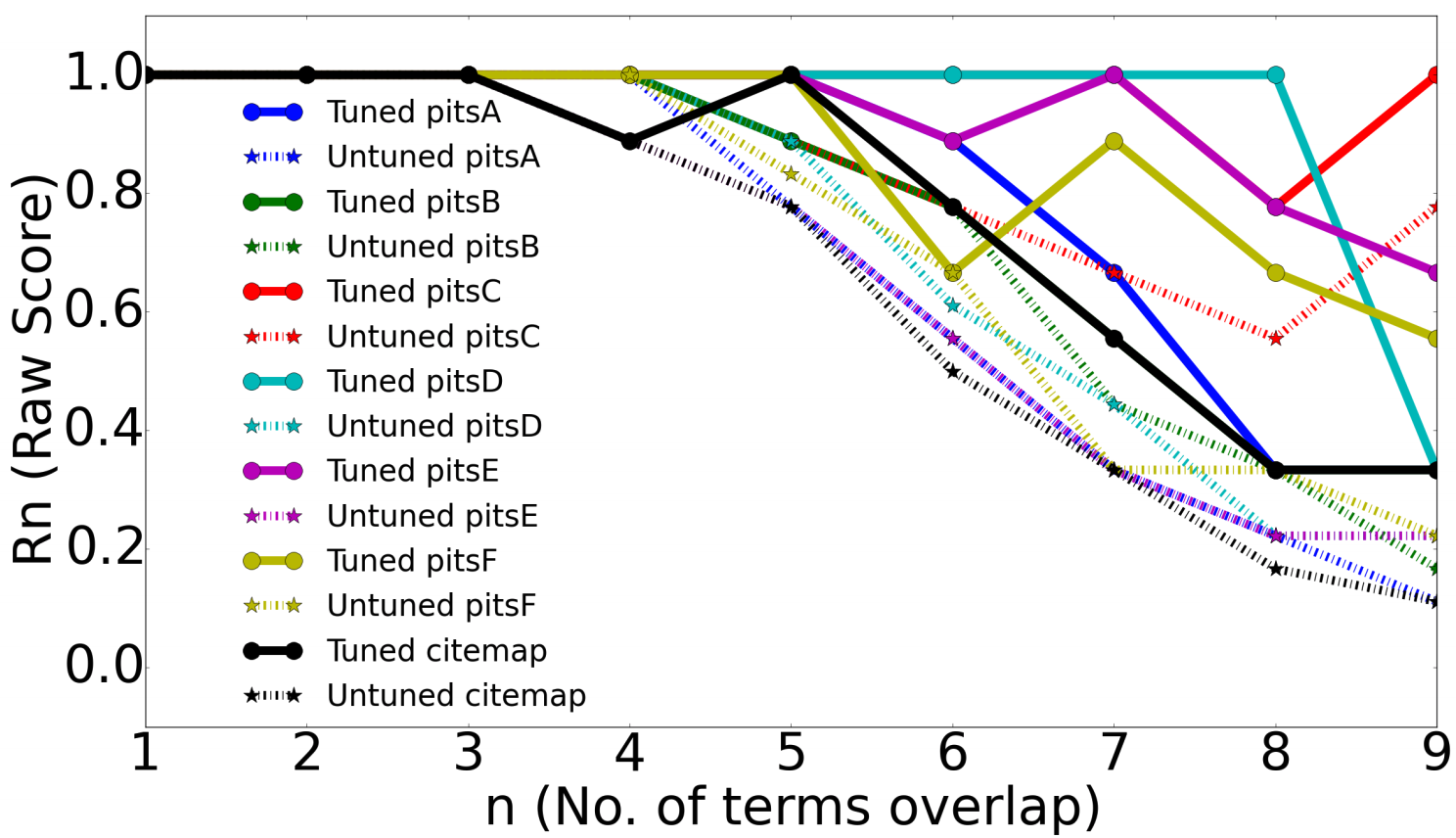}
  \footnotesize{{\bf Figure~\ref{fig:delta}a:}  {\em After} tuning: uses parameters learned by DE.}

        \includegraphics[width=0.9\linewidth]{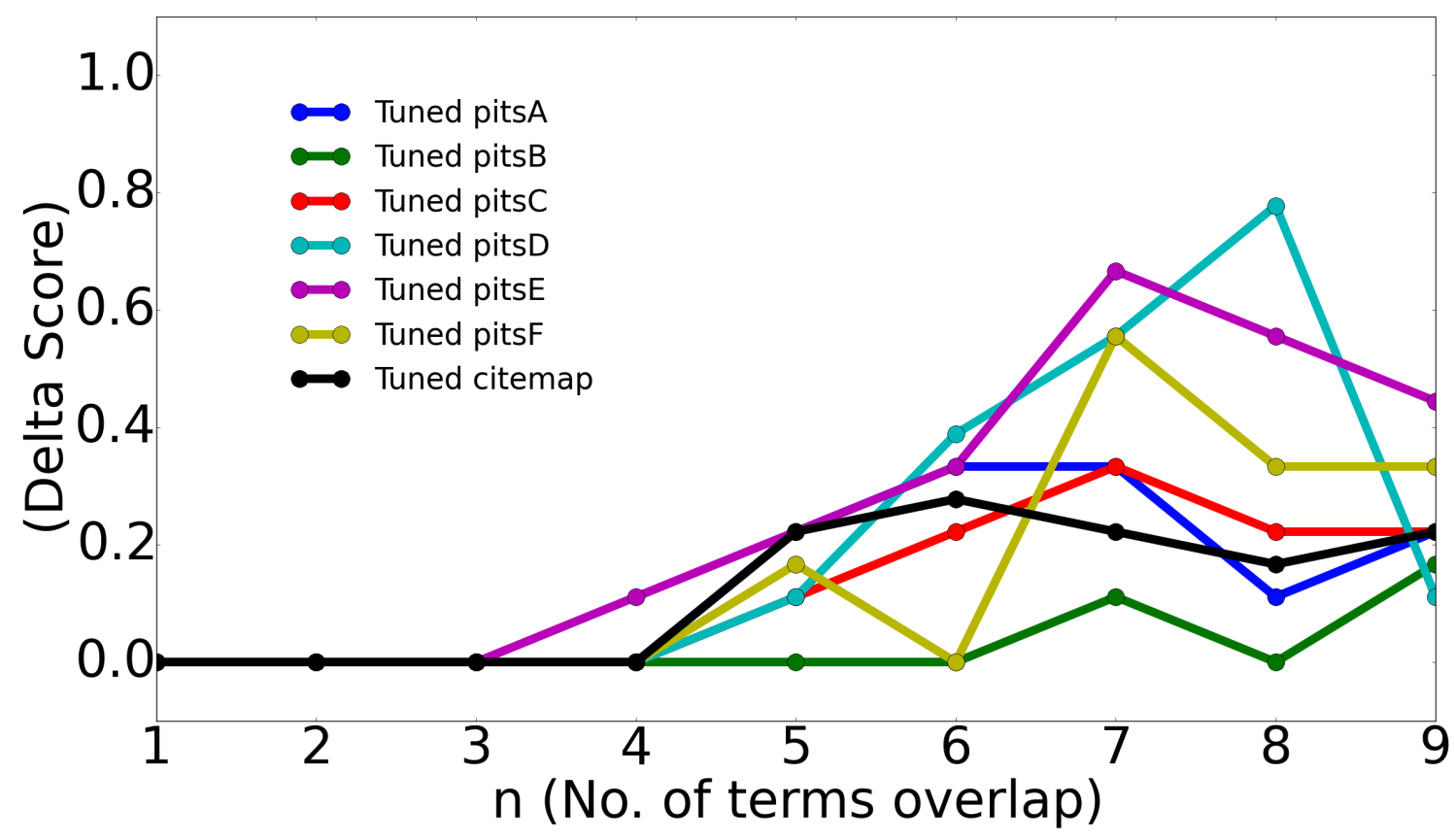}
  \footnotesize{{\bf Figure~\ref{fig:delta}b:}  {\em Delta = After - Before}.}
  \end{center}
    \caption{{\bf RQ1, RQ2} stability results over ten repeated runs. In these figures, {\em larger} numbers
    are {\em better}.}\label{fig:delta}
\end{figure}

 Figure~\ref{fig:delta}a and Figure~\ref{fig:delta}b show the stability improvement
 generated by tuning.
   Tuning never
  has any negative effect (reduces stability) and often has a large positive effect,
  particular  after 5 terms overlap.
   The largest improvement  we
   saw  in PitsD dataset which for up to 8 terms overlap was 100\% (i.e., was always
   found in all runs).
   Overall, after reporting topics of up to 7 words, in the majority case (66\%),
  those topics can be found in models generated using different input orderings.

  Continuing on the result of PitsD which achieves the highest improvement. We observed that  about 92\% of the data in that sample has the severity of level 3.  All the other Pits Datasets have mixed samples of severity level. When data is less skewed LDADE achieves the highest improvement which was not achieved when just LDA was being used. So, this makes it highly unimaginable to use just LDA for highly skewed data. This also emphasizes the use of LDADE more. 

  Since it is observed that LDADE achieves better stability, it gives us a chance to verify where the LDADE's recommended tunings can be most beneficial in SE task. For example,
  we performed the same analysis as we did in Table~\ref{tbl:olap} but this time
  using LDADE's recommended
  configurations.
  This leads to the 27   topics shown in Table~\ref{tbl:olap_stable}. Note that
  the topics
  found by our methods are far more stable
  than before:
  \bi
  \item The usual topic stability from LDADE is nearly 88\% (or more).
  \item
  At worst, the   topic stability  
  found by LDADE (55\%) is equal to the {\em
  median} instability of 
  Table~\ref{tbl:olap}.
  \item The worst stability of standard LDA in Table~\ref{tbl:olap} is over twice as bad as anything
  from  Table~\ref{tbl:olap_stable}.
  \ei

  Accordingly, our answer to {\bf RQ2} is:

\begin{lesson}
For stable clusters, tuning is strongly recommended for future LDA studies. $\alpha$ and $\beta$ matter the most for getting good clusters.
\end{lesson}

\subsection{\textbf{RQ3: Does LDADE improve text mining classification accuracy?}}\label{sect:rq3} 

We studied some other StackExchange websites data dump for classification results which were generated by Krishna et al \cite{krishna2016bigse}. These datasets are categorized into binary labels indicating which documents are relevant and non relevant. The goal of our DE was still to maximize the $\Re_n$ score. We did not change the goal of our DE based on classification task. After finding the optimal $k$, $\alpha$ and $\beta$, we trained a Linear Kernel SVM classifier using document topic distributions just like used by Blei et al~\cite{blei2003latent}. We used a 5-fold stratified cross validation to remove any sampling bias from the data. 80\% was used as training and 20\% as testing, and for the tuning phase, out of 80\%, 65\% was used for training and 15\% as validation set.

\begin{figure}[!htbp]
  \begin{center}
    \includegraphics[width=\linewidth]{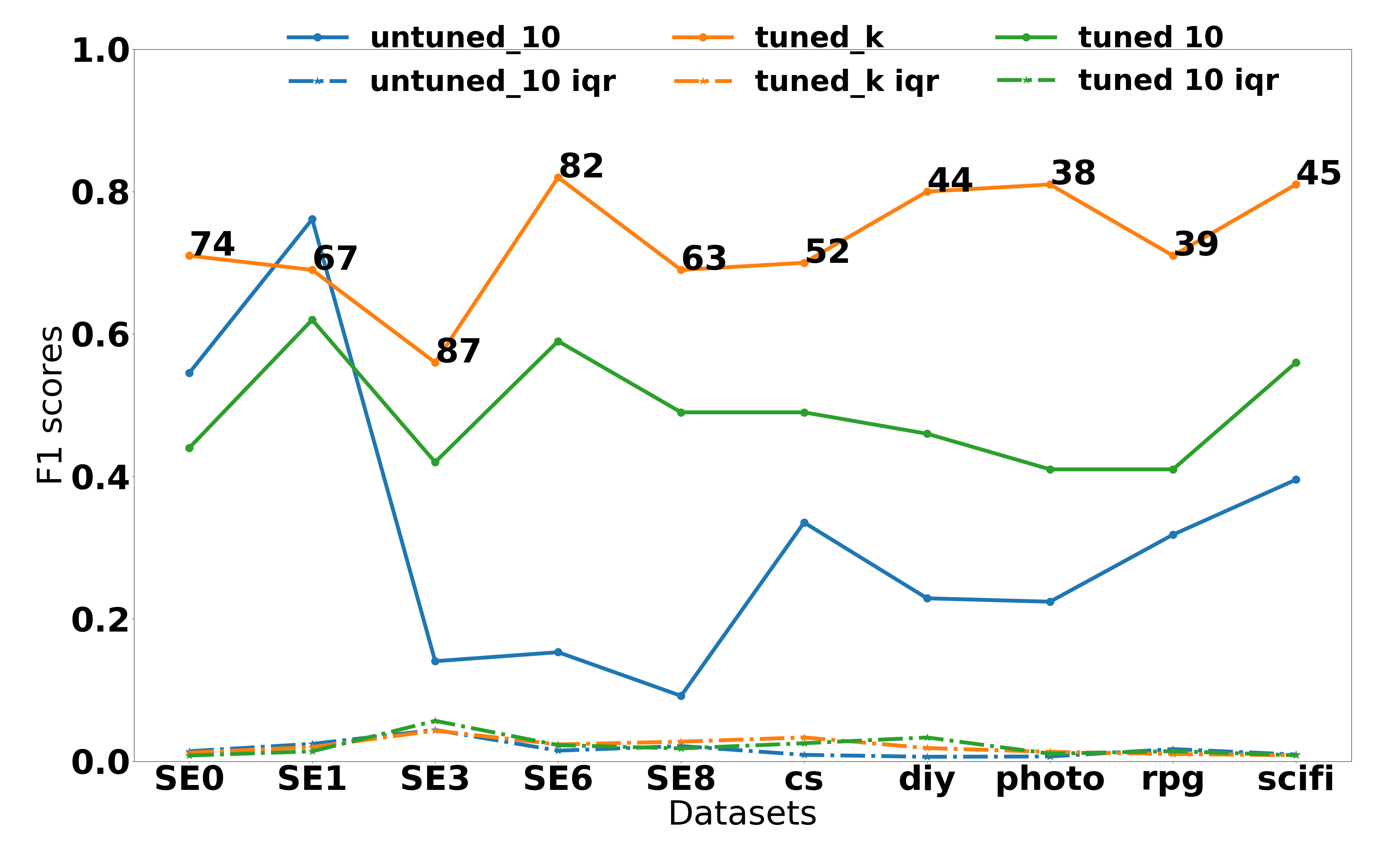}
    \end{center}
  \caption{Tuning and Untuned F1 results for Classification SE Task. The numbers in black show the $k$ values
  learned by LDADE.}\label{fig:classF1}  
\end{figure}

\begin{figure}[!htbp]
  \begin{center}
    \includegraphics[width=\linewidth]{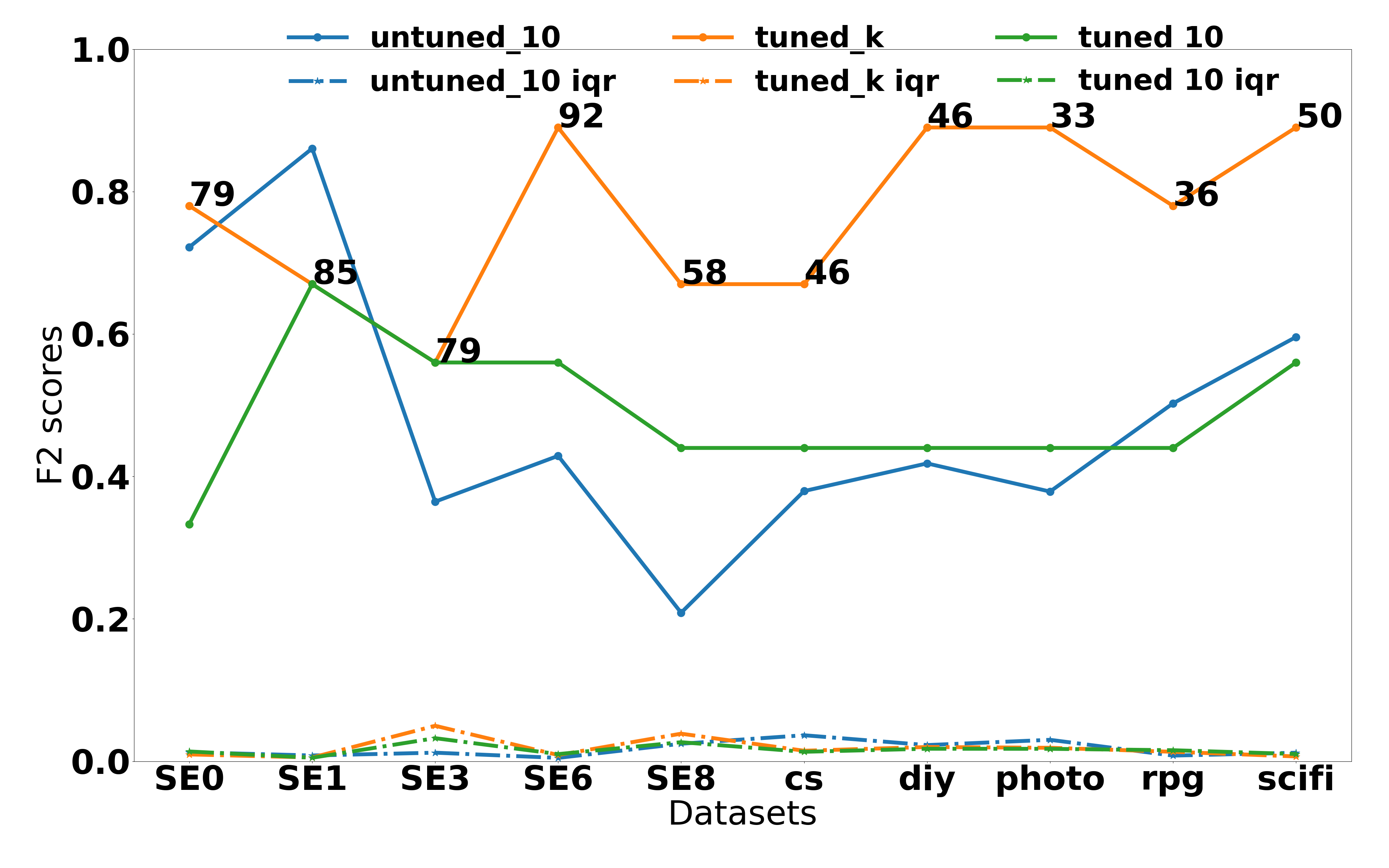}
    \end{center}
  \caption{Tuning and Untuned F2 results for Classification SE Task. The numbers in black show the $k$ values
  learned by LDADE.}\label{fig:classF2}  
\end{figure}

In Figures \ref{fig:classF1} and~\ref{fig:classF2}, the x-axis represents different datasets as generated. Y-axis represents the F1 score in Figure~\ref{fig:classF1} and in Figure~\ref{fig:classF2}, Y-axis represents F2 score
(from \eq{f})  which weights recall higher than precision~\cite{powers2011evaluation}. In the figures, ``untuned\_10'' indicates LDA used with default parameters of $k=10$, ``tuned\_k'' represents tuning of parameters selected from Table~\ref{tb:tuned} with value of $k$ written beside it, and ``tuned 10'' shows $k=10$ with $\alpha$ and $\beta$ tuned by our LDADE method. At the bottom of the figure, we show the variance which is inter-quartile (75th-25th percentile, i.e., IQR) range.

\begin{figure*}[!t]
    \centering
    \begin{minipage}{.33\textwidth}
        \captionsetup{justification=centering,singlelinecheck=off}
        \includegraphics[width=\linewidth]{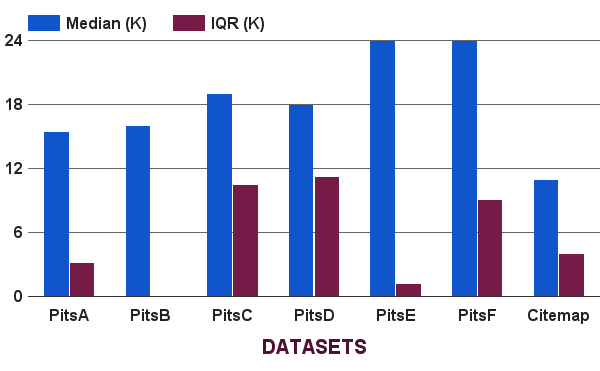}
        \caption{Datasets vs Parameter (k) variation}
        \label{RQ3:k}
    \end{minipage}%
    \begin{minipage}{.33\textwidth}
        \captionsetup{labelsep=space,justification=centering,singlelinecheck=off}
        \includegraphics[width=\linewidth]{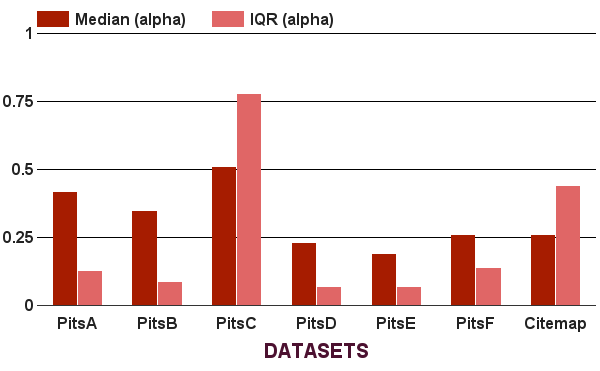}
        \caption{Datasets vs Parameter ($\alpha$) variation}
        \label{RQ3:a}
    \end{minipage}
    \begin{minipage}{.33\textwidth}
        \captionsetup{labelsep=space,justification=centering,singlelinecheck=off}
        \includegraphics[width=\linewidth]{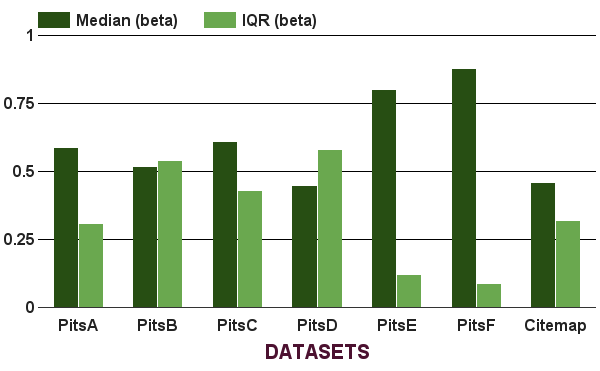}
        \caption{Datasets vs Parameter ($\beta$) variation}
        \label{RQ3:b}
    \end{minipage}
\end{figure*}

Just for completeness, we ran the Scott-Knot~\cite{jelihovschi2014scottknott} procedure for statistical differences using the A12 effect size~\cite{vargha2000critique}
test and bootstrapping~\cite{efron1994introduction} test which agreed that the division between tuned and untuned results were statistically significant (99\% confidence) and not a ``small'' effect ($A12 \ge 0.6$).

\noindent
Hence, we say:

\begin{lesson}
For any SE classification task, tuning is again highly recommended. And $k$ matters the most for a good classification accuracy.
\end{lesson}

\subsection{\textbf{RQ4: Do different data sets
      need different configurations to make LDA stable?}}
      \label{sect:diff}

Figures \ref{RQ3:k}, \ref{RQ3:a}, and \ref{RQ3:b} show the results of tuning for word overlap of $n=5$. Median values across 10 tunings and IQR (a non-parametric measure of variation
around the median value) are shown in these plots. Note that in Figure \ref{RQ3:k}, IQR=0 for  PitsB dataset which shows tuning always converged on the same final value.

These figures
show that the IQR ranges are being varied over median by about 50\% in most cases of datasets. . 
Some of the above numbers are far from the standard values, e.g., Garousi et al.~\cite{garousi2016citations} recommend using $k=67$ topics
yet in our data sets, best results were seen using $k \le 24$. These results suggest that how tuning selects the different ranges of parameters.
Clearly:

\begin{lesson}
  Do not  reuse tunings suggested by other researchers from other data sets.
  Instead, always re-tune for all new data.
\end{lesson}

\subsection{\textbf{RQ5}: \textbf{Are our findings consistent when using different kinds of LDA or with different implementations?}}
\label{sect:vem-gibbs}

To validate this research question, it was insightful to compare our results with: the Pits and Citemap results, executed in Scikit-Learn and Python running on a desktop machine as well as the Stackoverflow data set executed in Scala using Mllib running on a Spark cluster.

\begin{figure}[!b]
  \captionsetup{justification=centering}
  \includegraphics[width=\linewidth]{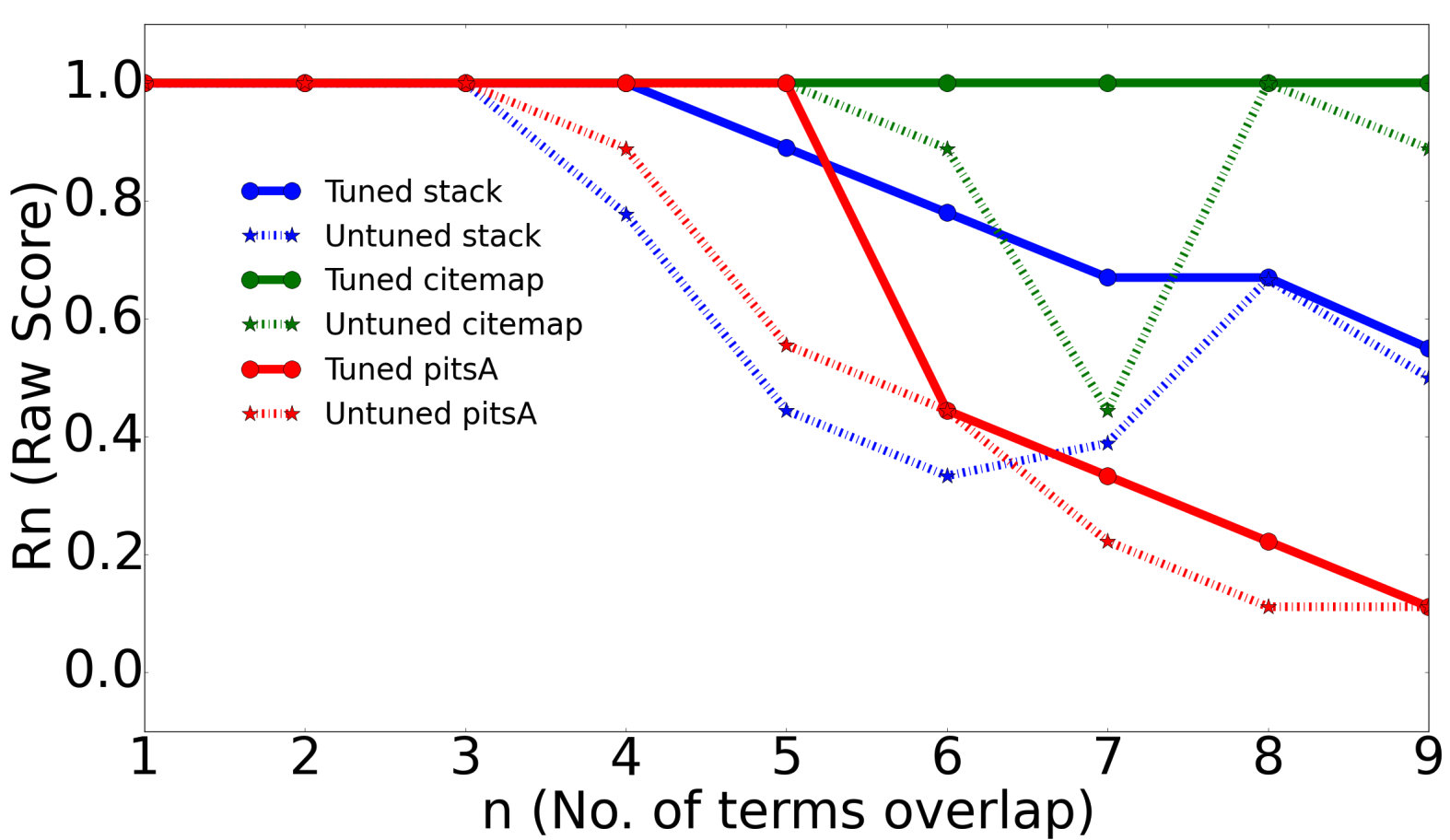}
  \caption{Spark Results}
  \label{python_spark}
\end{figure}

Figure \ref{python_spark} shows tuning results for Stackoverflow, Citemap, and PitsA 
   using Scala/Spark cluster (for results on other data sets, see https://goo.gl/UVaql1).
   
  Another useful comparison is to change the internal of the LDA, sometimes using VEM sampling and other times using Gibbs sampling.

\begin{figure}[!htbp]
  \captionsetup{justification=centering}
  \includegraphics[width=\linewidth]{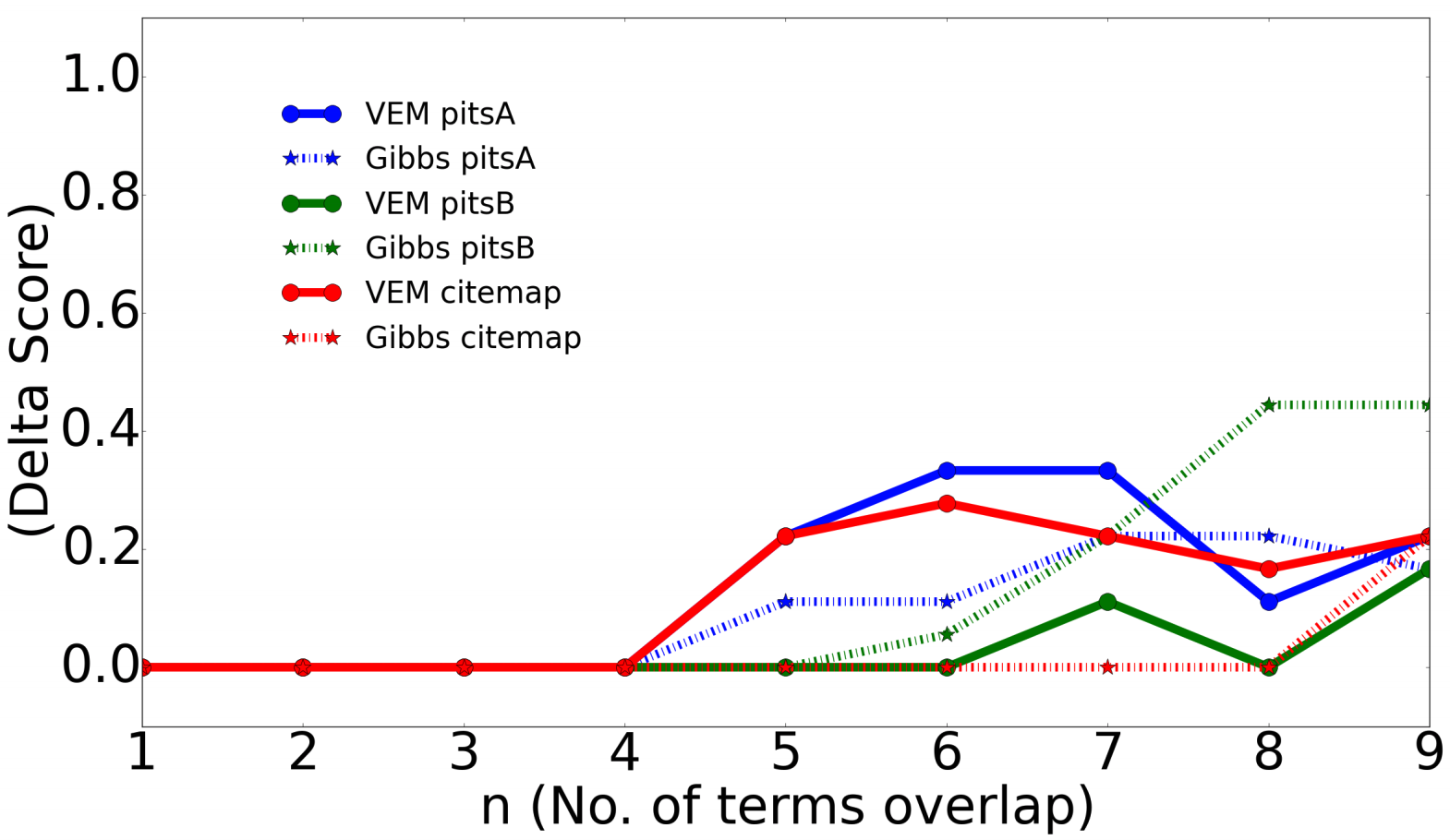}
  \caption{GIBBS vs VEM}
  \label{gibbs_vem}
\end{figure}

\begin{figure*}[!t]
    \centering
  \begin{minipage}{.49\textwidth}
        \captionsetup{labelsep=space,justification=centering}
        \includegraphics[width=\linewidth]{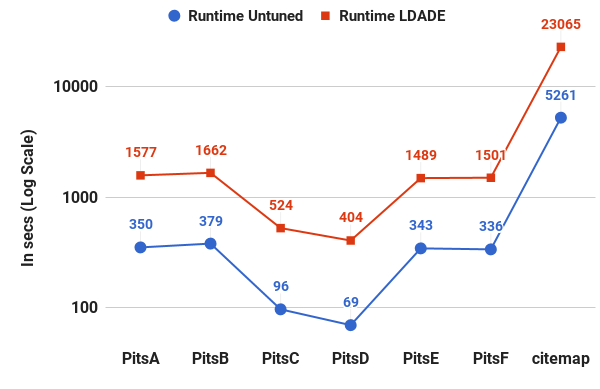}
  \caption{VEM: Datasets vs Runtimes}
  \label{RQ5 VEM}
  \end{minipage}
  \begin{minipage}{.49\textwidth}
        \captionsetup{justification=centering}
        \includegraphics[width=\linewidth]{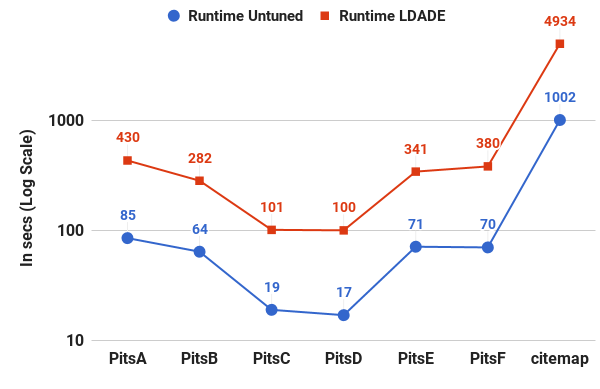}
  \caption{Gibbs: Datasets vs Runtimes}
  \label{RQ5 Gibbs}
    \end{minipage}%
    \vspace{-0.3cm}
\end{figure*}

\begin{figure}[!t]
  \includegraphics[width=\linewidth]{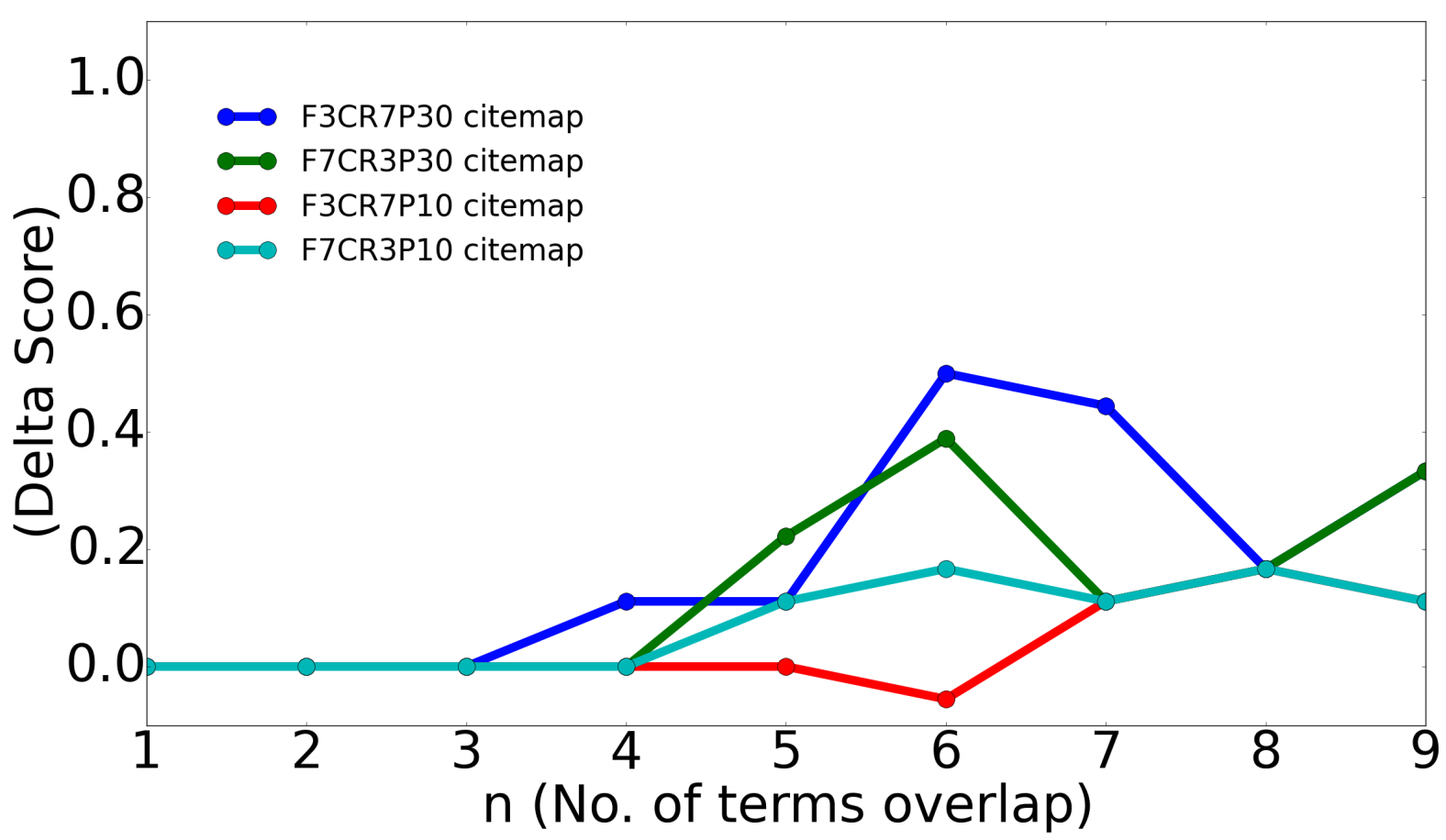}
  \caption{Terms vs Delta Improvement using fifferent settings of DE. The legends represent different values of $f$, $cr$ and $np$. So if legend says F3CR7P30, this shows $f=0.3$, $CR=0.7$ and $np=30$. Similarly for other legends}
  \label{fig:RQ4}
  \vspace{-0.3cm}
\end{figure}

  Figure~\ref{gibbs_vem} compares the  VEM vs Gibbs sampling (for results on other datasets, see https://goo.gl/faYAcg). When compared with the Python/desktop results of
   Figure~\ref{fig:delta} we see the same patterns that tuning never makes stability worse and sometimes, tuning dramatically improves it (in particular, see the Citemap results
   of  Figure~\ref{python_spark}).

   That said, there are some deltas between VEM and Gibbs where it seems tuning
   is more important for VEM than Gibbs (evidence: the improvements seen after
   tuning are largest for the  VEM results of  Figure~\ref{gibbs_vem} and at  https://goo.gl/faYAcg). The reason being, VEM only gets to a local optimum that depends on initialization and other factors in the optimization problem~\cite{asuncion2009smoothing}. On the other hand a good sampling algorithm like Gibbs sampling can achieve the global optimum. In practice because of finite number of samples that are used, sampling can give a sub-optimal solution. LDADE works as another kind of sampler which selects good priors at the initialization and can achieve better sub-optimal solutions, which is why LDADE has better benefits on VEM.

\begin{lesson}
  Instability is not due to any quirk in the implementation of LDA. Instability is consistent and LDADE can stabilize. 
\end{lesson}

\subsection{\textbf{RQ6: Is  tuning  easy?}}

The DE literature
recommends using a population size $np$ that is ten times larger than the number of parameters being
optimized~\cite{storn1997differential}.  For example, when tuning $k,\alpha$ and $\beta$,
the DE literature is recommending $np=30$.
Figure~\ref{fig:RQ4} explores $np=30$ vs the $np=10$ we use in Algorithm 2 (\fig{pseudo_DE})
(as well as some other variants of DE's $F$ and $CR$ parameters).
The figure shows results just for Citemap and, for space reasons, results
relating to other data sets are shown at https://goo.gl/HQNASF.
After reviewing the results from all the datasets, we can say that there is not much of an improvement by using different $F$, $CR$, and Population size. So our all other experiments used $F=0.7$, $CR=0.3$ and $np = 10$.
Also:

\begin{lesson}
  Finding stable parameters for
  topic models is easier than standard optimization tasks.
\end{lesson}

\subsection{\textbf{RQ7: Is tuning extremely slow?}}

Search-based SE methods can be very slow. Wang et al.~\cite{wang2013searching} once needed 15
years of CPU time to find and verify the tunings required for software
clone detectors. Sayyad et al.~\cite{sayyad2013scalable} routinely used
$10^6$ evaluations (or more) of their models in order to extract
products from highly constrained product
lines. Hence, before recommending any
search-based method, it is wise to consider the runtime cost of that
recommendation.

To understand our timing results, recall that untuned and LDADE use
Algorithm~1 and Algorithm~2 respectively. Based on the psuedocode
shown above, our pre-experimental theory is that
tuning will be three times slower than not tuning (since DE is taking 3 generations to terminate).
 
Figures \ref{RQ5 VEM} and \ref{RQ5 Gibbs} check if this theory
holds true in practice. Shown in blue and red are the
  runtimes required to run LDA untuned and LDADE (respectively).  The
  longer runtimes (in red) include the times required for DE to find
  the tunings. Overall, tuning slows down LDA by a factor of up to
  five (which is very close to our theoretical prediction).
  Hence, we say:

  \begin{lesson}
    Theoretically and empirically, LDADE costs three to five times more runtime
    as much as using untuned LDA.
  \end{lesson}

  While this is definitely more than not using DE, but this may not be an arduous increase
  given modern cloud computing environments.

\subsection{\textbf{RQ8: How better LDADE is compared against other methods for tuning LDA?}}
\label{sect:rq8}

In this section,
we perform two comparisons of LDADE  against    prior work LDA-GA~\cite{panichella2013effectively} and random search.

In the first comparison against LDA-GA, we argue that Panichella et al.~\cite{panichella2013effectively} did not consider order effects, which could result in instability of LDA when they proposed their method of LDA-GA. 
Another issue could be with their use of Genetic Algorithms to find optimal configurations. Such GAs can be very time consuming to run.
 Fu et al.~\cite{fu2017easy} argues for faster software analytics,
 saying that the saved CPU time
 could be put to better uses.

\begin{figure*}[!t]
    \centering
  \begin{minipage}{.49\textwidth}
        \captionsetup{labelsep=space,justification=centering}
       \includegraphics[width=\linewidth]{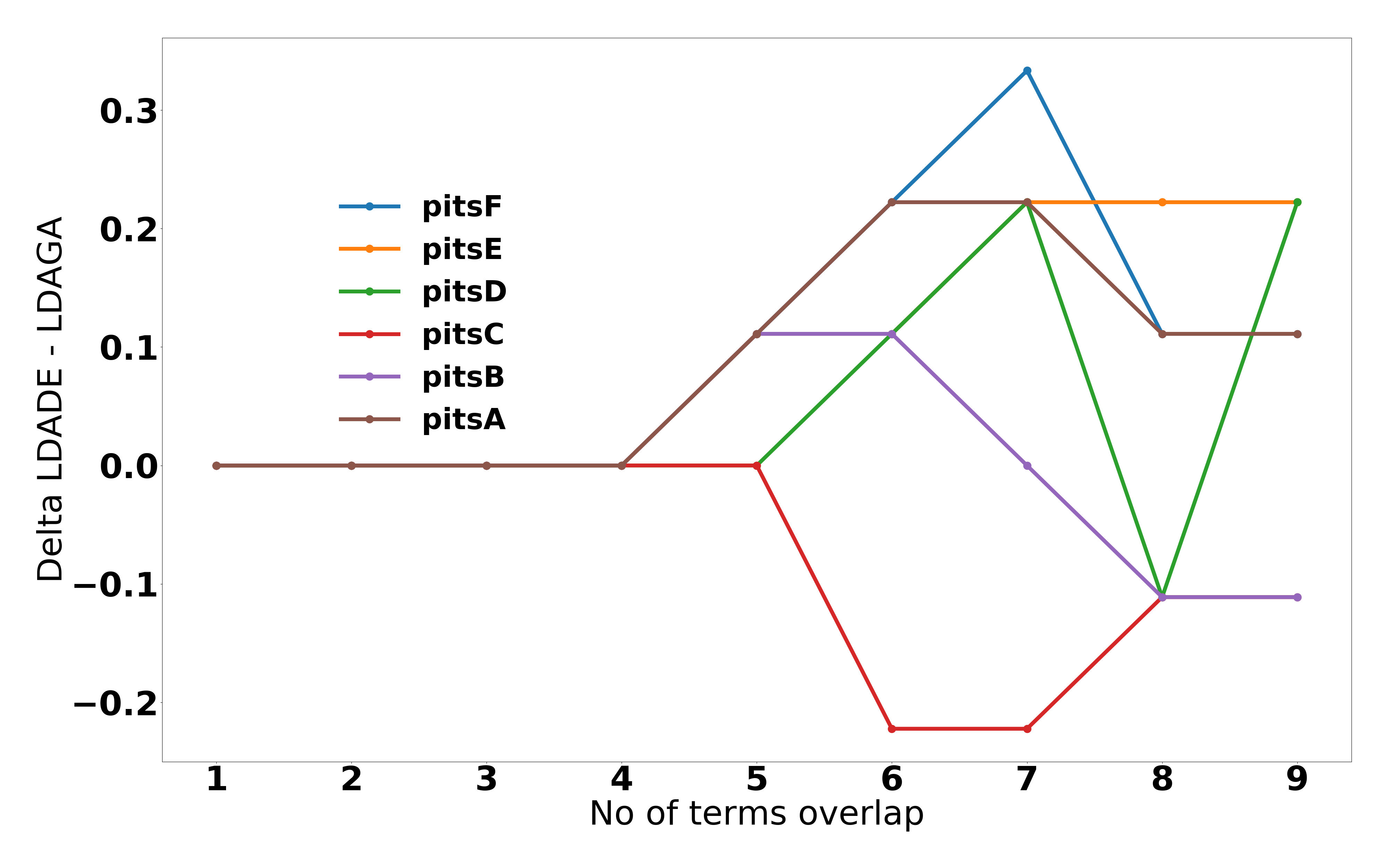}
  \caption{Terms vs Delta comparison of LDADE against LDA-GA}
  \label{fig:ldaga}
  \end{minipage}
  \begin{minipage}{.49\textwidth}
        \captionsetup{justification=centering}
        \includegraphics[width=\linewidth]{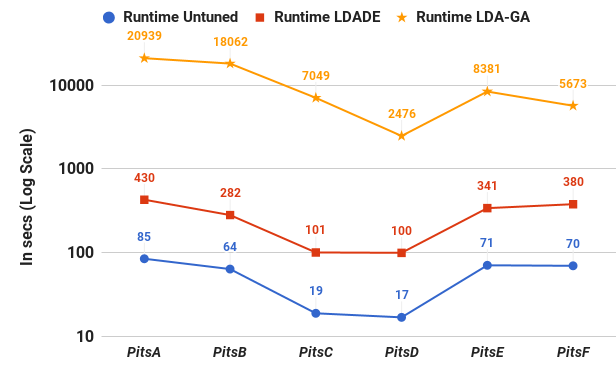}
  \caption{Gibbs: Runtime comparisons between LDADE and LDA-GA}
  \label{fig:run_ldaga}
    \end{minipage}%
\end{figure*}

We ran LDA-GA with their same settings but on our datasets and report the delta of $\Re_n$ of LDADE against $\Re_n$ of LDA-GA. In Figure~\ref{fig:ldaga}, the positive value shows LDADE performed better than LDA-GA and negative shows LDA-GA performed better. In only about 2 cases (PitsB, PitsC), it is seen that LDA-GA performed better. The reason being, LDA-GA takes a larger number of evaluations which was able to find a better sub-optimal configuration, but it took somewhere between 24-70 times more time than LDADE (See Figure~\ref{fig:run_ldaga}). In Figure~\ref{fig:run_ldaga}, we are seeing variations in the difference between the runtimes of LDADE and LDA-GA in these datasets. This is due to the settings of GA. LDA-GA has early termination criteria which depends on the quality of dataset. LDA-GA should take somewhere from 1000 to 10,000 evaluations where as our DE would take about 30 evaluations. If a particular dataset is less skewed, it will be terminated early in LDA-GA. These results verify that our initial assumptions hold true, i.e., we can get much faster and more stabler results with LDADE than LDA-GA.

\begin{figure*}[!t]
    \centering
  \begin{minipage}{.49\textwidth}
        \captionsetup{labelsep=space,justification=centering}
        \includegraphics[width=\linewidth]{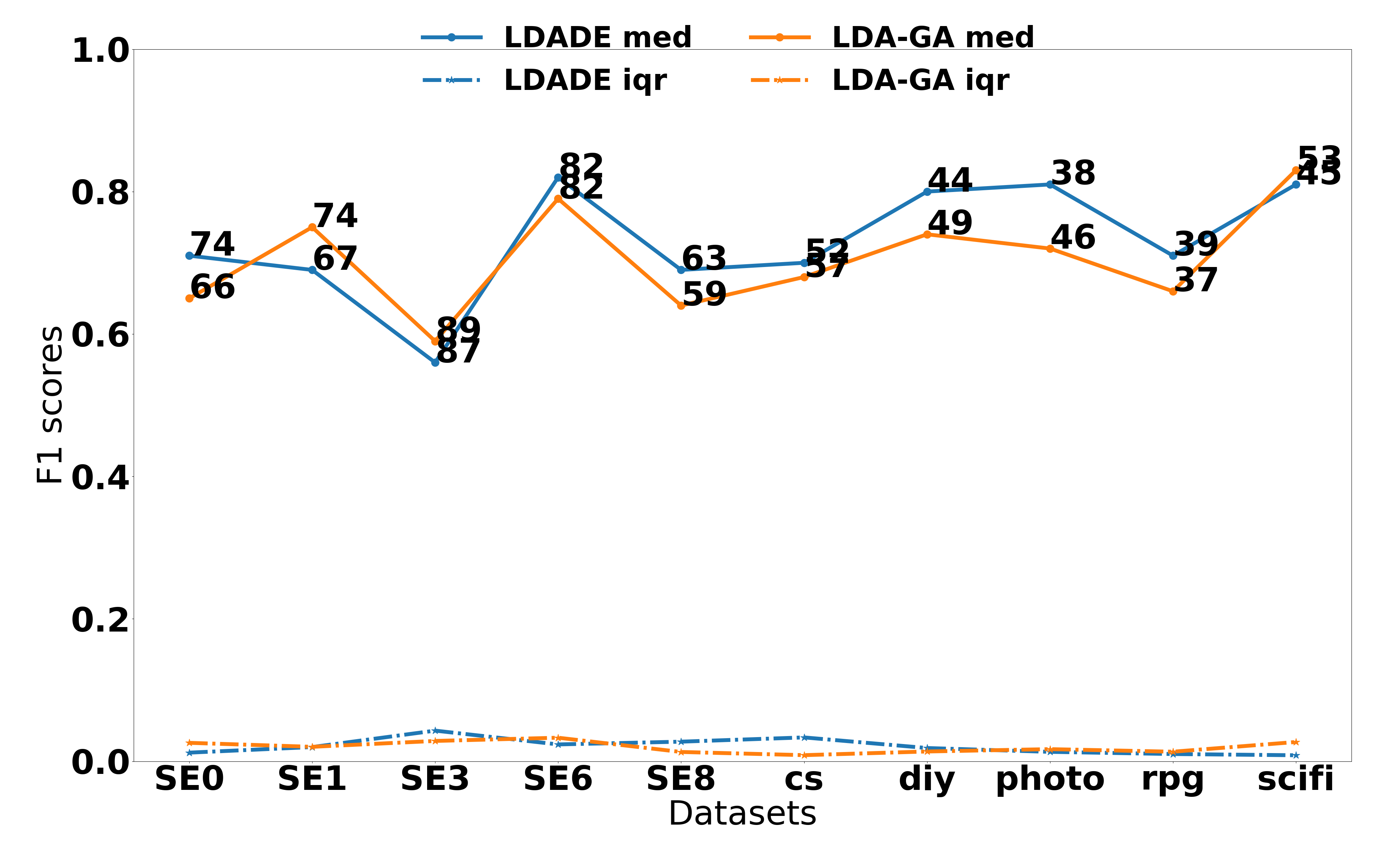}
  \caption{Comparison of LDADE and LDA-GA for Classification SE Task. The numbers in black show the $k$ values
  learned by individual methods.}
  \label{fig:ldade_ga}
  \end{minipage}
  \begin{minipage}{.49\textwidth}
        \captionsetup{justification=centering}
        \includegraphics[width=\linewidth]{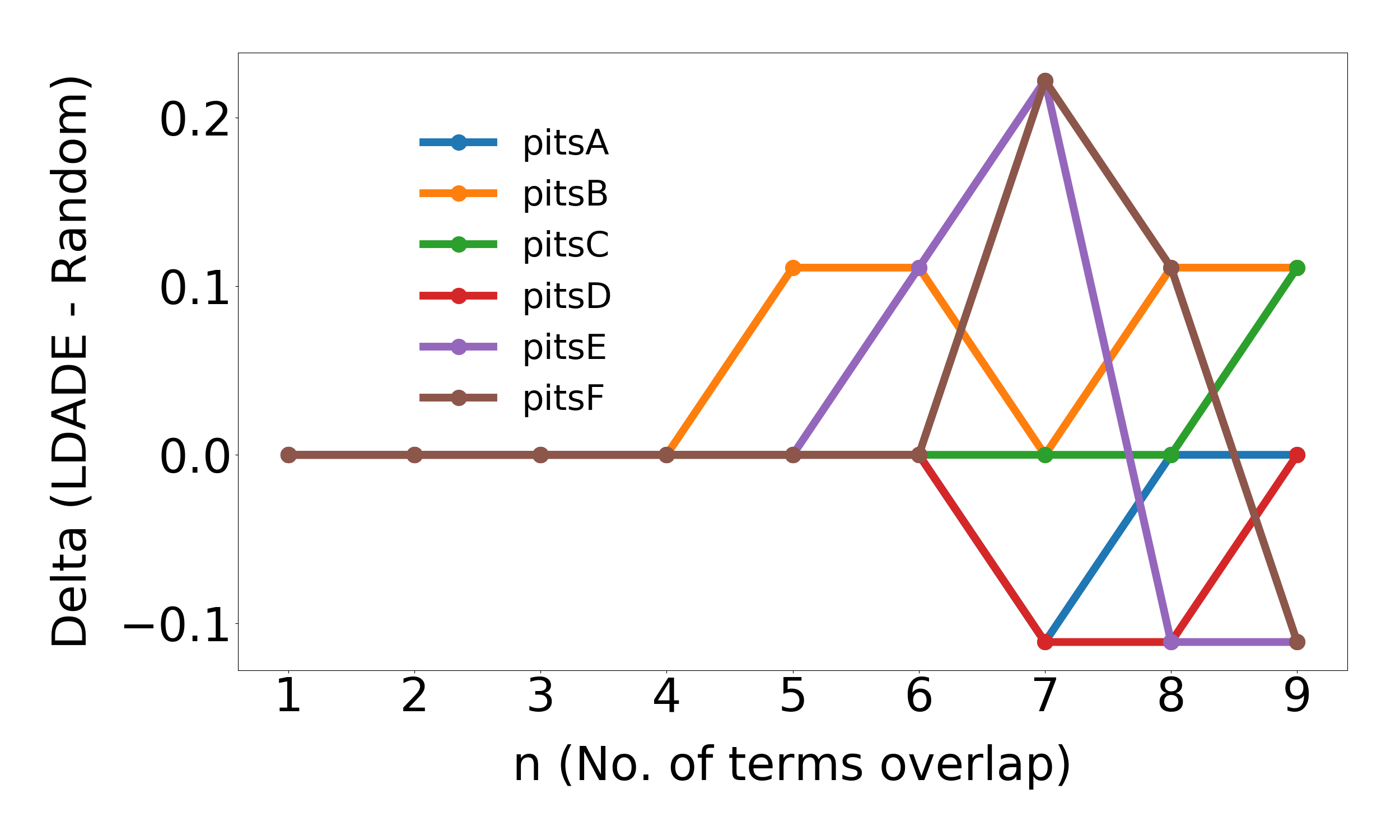}
  \caption{Terms vs Delta comparison of LDADE against Random Search. Higher the improvement, better is LDADE.}
  \label{fig:random}
    \end{minipage}%
    
\end{figure*}

We further extend our comparison of LDADE against LDA-GA for the classification task and details are the same as described in Section~\ref{sect:rq3}.
 From Figure~\ref{fig:ldade_ga}, we   see  that in
 7/10 of our datasets, LDADE performs somewhat bigger.
 That said,
 the overall trend is that these two methods have similar F1 performances on different datasets
(the reason being that   $k$ value found by both methods is very similar).

Overall, we recommend LDADE since:
\bi
\item Figure~\ref{fig:ldade_ga}
tell us that the
median classification performance of
  LDADE is slightly better than LDA-GA.
\item  Figure~\ref{fig:run_ldaga} tells us that
LDADE runs  24-70 times faster.
\item  
And Figure~\ref{fig:ldaga} tells us
that LDADE produces stabler
topics.
\ei
 
Secondly, we performed the comparison of LDADE against random search procedure.
We performed random search with 30 evaluations to find different configurations of LDA (since DE took 30 evaluations mentioned in Section~\ref{sect:tuning}) and observed that the improvement achieved with LDADE outperforms the improvement achieved with random search (Figure~\ref{fig:random}). DE is smarter than random search because of 2 reasons:
\bi
\item It samples from a space that is continuously improving. In every generation, after item $i$ the frontier contains   items
  better than at least one thing
seen in the last generation (as do  items $1..i$).   During the next generation, mutation happens between 3 better candidates
and, even after being 50\% through one generation,
odds are that the 3 candidates have all passed the ``better''
test. So DE builds better solutions
from a space of candidates that it is continually
refining and improving
\item GAs, SA and others mutate all their
attributes independently. But DE  supports vector-level mutation that retain the association between variables in the space~\cite{das2011differential}.
\ei

  \begin{lesson}
    LDA-GA is afflicted  by order effects and it is slower than LDADE.  LDADE also achieves stabler results than LDA-GA and random search. 
  \end{lesson}

\subsection{\textbf{RQ9: Should topic modeling be used ``off-the-shelf'' with their default tunings?}}
\label{sect:rq9}

  Figure~\ref{fig:delta} shows that there is much benefit in tuning.
  Figures \ref{RQ3:k}, \ref{RQ3:a}, and \ref{RQ3:b} show that
  the range of ``best'' tunings is very dataset specific. 
    Based on the above, we assert that
  for a new dataset,
  the off-the-shelf tunings
  may often fall far from the useful range.
  Figures \ref{RQ5 VEM} and \ref{RQ5 Gibbs} show that tuning is definitely
  slower than otherwise, but the overall cost is not prohibitive.
  Hence:
  \begin{lesson}
    Whatever the goal is, whether using the learned topics, or cluster distribution for classification
    we cannot recommend using ``off-the-shelf'' LDA.
  \end{lesson}

\section{Threats to Validity}
\label{sect:validity}

As with any empirical study, biases can affect the final
results. Therefore, any conclusions made from this work must be considered with the following issues in mind:

\textbf{\textit{Sampling bias}} threatens any experiment, i.e., what matters there may not be true here. For example,
the data sets used here come after various pre-processing steps and could change if pre-processed differently. And that is why, all our datasets can be downloaded from the footnotes of this paper and researchers can explore further. Even though we used so many data sets, there could be other datasets for which our results could be wrong or have lesser improvement.

\textbf{\textit{Learner bias}}: For running LDA, we selected other parameters as default which are of not much importance. But there could be some datasets where by tuning them there could be much larger improvement. And for RQ2, we only experimented with linear kernel SVM. There could be other classifiers which can change our conclusions. Data Mining is a large and active field and any single study can only use a small subset of the known data miners.

\textbf{\textit{Evaluation bias}}: This paper uses topic similarity ($\Re_n$) and F2 measures of evaluation but there are other measures which are used in software engineering which
includes perplexity, performance, accuracy, etc. Assessing
the performance of stable LDA is a clear direction for future work.

\textbf{\textit{Order bias}}: With each dataset, how data samples are picked and put into LDA is completely random. Since this paper also consider input order effects, though there could be times when the input order could be with lesser variance. To mitigate this order bias, we ran the experiment 10 times by randomly changing the order of the data samples each time.

Another threat to validity of this work is that it is a quirk of the control
parameters used within our DE optimizer.
We have some evidence that this is not the case.
Figure~\ref{fig:RQ4} and other results\footnote{https://goo.gl/HQNASF} explored a range of DE tunings and found
little difference across that range. Also, Table~V explores another choice within DE -- how
many evaluations to execute before terminating DEs. All the results in this paper use an
evaluation budget of 30 evaluations. Table~V
compares results across different numbers of evaluations. While clearly,
the more evaluations the better, there is little improvement after the 30 evaluations used in this paper. 



We also acknowledge that for now we validated the improvement of LDADE over LDA in an unsupervised task (see Table~\ref{tbl:olap} and ~\ref{tbl:olap_stable}) and in one case study of supervised task. The gains shown in this study may not be prominent when tested on any other unsupervised or supervised task. In future work,
we need to
assess the advantage of  LDADE for other SE tasks.

\begin{table}[!t]
\scriptsize
\begin{center}
\caption{Evaluations vs Stability Scores}
\label{tb:tablename1}
\begin{tabular}{|c|c|c|c|c|}
\hline 
\textbf{Datasets\textbackslash Evaluations} & \textbf{10} & \textbf{20} & \textbf{30} &
\textbf{50} \\[0.5ex]
\hline
PitsA & 0.9 & 0.9 & 1.0 & 1.0\\ 
\hline
PitsB & 0.9 & 0.9 & 0.9 & 1.0 \\
\hline
PitsC & 0.9 & 1.0 & 1.0 & 1.0\\ 
\hline
PitsD & 0.9 & 1.0 & 1.0 & 1.0\\ 
\hline
PitsE & 0.9 & 0.9 & 1.0 & 1.0\\
\hline
PitsF & 0.9 & 0.9 & 0.9 & 0.9\\
\hline
Citemap & 0.67 & 0.67 & 0.77 & 0.77\\
\hline
Stackoverflow & 0.6 & 0.7 & 0.8 & 0.8\\
\hline
\end{tabular}
\end{center}
\end{table}

The conclusions of this paper are based on a finite number of data sets and it is possible
that other data might invalidate our conclusions. As with all analytics papers, any researcher can do is to make their conclusions and materials public, then encourage
other researchers to repeat/refute/improve their conclusions.

\section{Conclusion}

Based on the above, we offer some general recommendations. Any study that shows the topics learned from LDA, and uses them to make a particular
conclusion, needs to first tune LDA. We say this since the topics learned from untuned LDA are unstable, i.e., different input orderings will lead to different conclusions. However, after tuning, stability can be greatly increased.

Unlike the advise of Lukins et al.~\cite{lukins2010bug}, LDA topics should not be reported as the top ten words.
  Due to order effects, such a report can be highly incorrect.
  Our results show that up to eight words can be reliably reported, but only
  after tuning for stability using tools like LDADE.

Any other studies which are making use of these topic distributions need to be tuned first before using them in their further tasks. We do not recommend to use someone else's pre-tuned LDA since, as shown in this study,  the best LDA tunings vary from data set to data set. These results also ask us to revisit the previous case studies which have used LDA using ``off-the-shelf'' parameters to rerun by tuning these parameters using automated tools like LDADE. Our experience is that this recommendation is not an arduous demand since tuning adds less than a factor of five to the total run times of an LDA study.

More generally, we comment that the field of software analytics needs to make far more use of search-based software engineering in order
to tune their learners. In other work, Fu et al. have shown that tuning significantly helps defect prediction~\cite{fu2016tuning} and an improvement also shown for LDA~\cite{panichella2013effectively}. In this work, we have shown that tuning significantly helps topic modeling by mitigating a systematic error in LDA  (order effects that lead to unstable topics). The implication of this study for other software analytics tasks is now an open
and pressing issue. 
In how many domains can search-based SE dramatically improve software analytics?

\section*{Acknowledgements}
		The work is partially funded by NSF award \#1506586.
	
\balance
\bibliographystyle{elsarticle-num}
\medskip

\end{document}